\documentclass[copyright,creativecommons]{eptcs}
\usepackage{etex}
\providecommand{\event}{VPT 2019} % Name of the event you are submitting to
\usepackage{underscore}           % Only needed if you use pdflatex.

\usepackage[T1]{fontenc}
\usepackage[english]{babel}
\usepackage[utf8]{inputenc}
\usepackage[scaled]{}
\usepackage{cite}
\usepackage{listings}
\usepackage{latexsym, amssymb, amsmath, amsfonts, yfonts, mathrsfs, amsthm}
\usepackage{syntax}
\usepackage{mathtools}
\usepackage{multicol}
\usepackage{tikz-cd}
\usepackage{color}
\usepackage{graphicx}
\usepackage{graphics}
\usepackage{listings}
\usepackage{import}
\usepackage{stmaryrd}
\usepackage{galois}
\usepackage{array}
\usepackage{pdflscape}
\usepackage[ruled, linesnumbered]{algorithm2e}
\usepackage{hyperref}
\usepackage{capt-of}
\usepackage{makecell}
\usepackage{framed}
\usepackage{mdwlist}
\usepackage{float}

\usepackage{graphicx}
\usepackage[labelformat = empty]{subcaption}
\usepackage[font=small,skip=4pt]{caption}

\usepackage{pgf}
\usepackage{tikz, wrapfig,fancybox}
\usetikzlibrary{arrows,automata}
\usepackage[export]{adjustbox}

\makeatletter
\newenvironment{CenteredBox}{% 
	\begin{Sbox}}{% Save the content in a box
	\end{Sbox}\centerline{\parbox{\wd\@Sbox}{\TheSbox}}}% And output it centered
\makeatother

\newtheorem{theorem}{Theorem}

\newtheorem{definition}{Definition}
\newtheorem{example}{Example}

\SetKwInput{KwData}{Input}
\SetKwInput{KwResult}{Output}

\lstdefinelanguage{JS}
{ keywords={var, function, typeof, this, undefined, parseInt},
	otherkeywords={},
	basicstyle=\fontsize{10pt}{12pt}\selectfont\ttfamily,
	keywordstyle=\bfseries\color{blue},
	sensitive=false,
	commentstyle=\color{purple!40!black},
	showspaces=false,
	tabsize=1,
	literate= {~}{\texttt{\phantom{m}}}1 {`}{$\texttt{\%}$}1 {?}{$\texttt{\$}$}1, %{~}{$\sim$}{1} 
	showstringspaces=false,emph={3}{\tiny }
	showtabs=true,
	morecomment=[l]{//},
	morecomment=[s]{/*}{*/},
	morestring=[b],
	breaklines=true,
	breakindent=12pt
}

\definecolor{dkred}{rgb}{.6,0,0}
\definecolor{dkgreen}{rgb}{0,.5,0}
\definecolor{dkblue}{rgb}{0,0,.6}
\definecolor{dkyellow}{cmyk}{0,0,.8,.3}
\definecolor{lightgray}{rgb}{.95,.95,.95}
\definecolor{darkgray}{rgb}{.3,.3,.3}
\definecolor{darkblue}{rgb}{0,0,.20}
\definecolor{purple}{rgb}{0.65, 0.12, 0.82}

\newif\ifdraft\draftfalse

\newcommand{\xremark}[2]{{\color{dkgreen}(#1: #2)}}
\newcommand{\yremark}[2]{{\color{dkred}(#1: #2)}}
\newcommand{\zremark}[1]{{\color{dkyellow}#1}}

\ifdraft
\newcommand{\va}[1]{\yremark{VA}{#1}}
\newcommand{\im}[1]{\xremark{IM}{#1}}
\newcommand{\added}[1]{\zremark{#1}}
\else
\newcommand{\va}[1]{}
\newcommand{\im}[1]{}
\newcommand{\added}[1]{}
\fi

\newcommand{\sset}[2]{\{\;#1 \;|\; #2 \;\}}

\newcommand{\fa}{\mathsf{\DFA_{/\equiv}}}
\newcommand{\DFA}{\mbox{\sc Dfa}}
\newcommand{\leqfa}{\sqsubseteq_{\mbox{\tiny \DFA}}}

\newcommand{\lubfa}{\sqcup_{\mbox{\tiny \DFA}}}
\newcommand{\glbfa}{\sqcap_{\mbox{\tiny \DFA}}}

\newcommand{\latticefa}{\langle \fa, \leqfa, \lubfa, \glbfa,\minimize(\varnothing),\minimize(\Sigma^*)\rangle}

\newcommand{\rightquotient}[2]{\mbox{\sc Rq}(#1, #2)}
\newcommand{\pref}[1]{\mbox{\sc Pr}(#1)}
\newcommand{\suff}[1]{\mbox{\sc Su}(#1)}
\newcommand{\isuff}[2]{\mbox{\sc Su}(#1,#2)}

\newcommand{\minimize}{\mathsf{Min}}
\newcommand{\lang}{\mathscr{L}}
\newcommand{\aut}{\mbox{\tt A}}
\newcommand{\lin}{\mbox{\tt L}}

\newcommand{\mujs}{\mathsf{IMP}}
\newcommand{\mktype}[1]{\mbox{\scshape{#1}}}
\newcommand{\cval}{\mathbb{V}}
\newcommand{\cbool}{\mathbb{B}}
\newcommand{\cint}{\mathbb{Z}}
\newcommand{\cstr}{\mathbb{S}}
\newcommand{\cnan}{\code{NaN}}
\newcommand{\expr}{\mbox{\bf Exp}}
\newcommand{\stmt}{\mbox{\bf Stmt}}
\newcommand{\block}{\mbox{\bf Block}}
\newcommand{\id}{\mbox{\bf Id}}
\newcommand{\true}{\mbox{\tt true}}
\newcommand{\false}{\mbox{\tt false}}

\newcommand{\substring}[2]{\mbox{\tt substring(}#1\mbox{\tt ,}#2\mbox{\tt )}}
\newcommand{\substringf}[3]{\mbox{\sl substring(}#1\mbox{\sl ,}#2\mbox{\sl ,}#3\mbox{\sl )}}

\newcommand{\charat}[1]{\mbox{\tt charAt(}#1\mbox{\tt)}}
\newcommand{\charata}[2]{\mathsf{CA}^\sharp(#1,#2)}

\newcommand{\length}{\mbox{\tt length}}

\newcommand{\lengtha}[1]{\mathsf{LE}^\sharp(#1)}

\newcommand{\indexof}[1]{\mbox{\tt indexOf(}#1\mbox{\tt )}}
\newcommand{\indexofa}[2]{\mathsf{IO}^{\sharp}(#1,#2)}

\newcommand{\conca}[2]{\mathsf{CC}^\sharp(#1,#2)}
\newcommand{\concs}[2]{\mbox{{\sc Cc}}(#1,#2)}

\newcommand{\code}[1]{\mbox{\lstinline@#1@}}

\newcommand{\semc}[1]{\llbracket \mbox{\lstinline@#1@}\rrbracket}

\newcommand{\sem}[1]{\llbracket #1\rrbracket}

\newcommand{\asemc}[1]{\llbracket \mbox{\lstinline@#1@}\rrbracket^\sharp}

\newcommand{\asem}[1]{\llbracket #1\rrbracket^\sharp}

\newcommand{\states}{\mktype{States}}

\newcommand{\aval}{\cval^\sharp}
\newcommand{\abool}{\cbool^\sharp}
\newcommand{\aint}{\cint^\sharp}
\newcommand{\astr}{\cstr^\sharp}
\newcommand{\anan}{\cnan}
\newcommand{\interval}{\mathsf{Int}}
\newcommand{\nint}{\mathbb{Z}}

\newcommand{\substringl}[3]{\mbox{\sc Ss}(#1,#2,#3)}
\newcommand{\substringa}[3]{\mathsf{SS}(#1,#2,#3)}

\newcommand{\factors}[1]{\mbox{\sc Fa}(#1)}
\newcommand{\charats}[2]{\mbox{\sc Ca}(#1,#2)}
\newcommand{\indexofs}[2]{\mbox{\sc Io}(#1,#2)}
\newcommand{\lengths}[1]{\mbox{\sc Le}(#1)}
\newcommand{\substringlright}[3]{\mbox{\sc Ss}^\rightarrow(#1,#2,#3)}

\newcommand{\substringaright}[3]{\mathsf{SS}^\rightarrow(#1,#2,#3)}

\newcommand{\substringalr}[2]{\mathsf{SS}^\leftrightarrow(#1,#2)}
\newcommand{\substringlr}[2]{\mbox{\sc Ss}^\leftrightarrow(#1,#2)}
\newcommand{\asubstring}[3]{\mathsf{SS}^\sharp(#1,#2,#3)}

\newcommand{\asuff}[1]{\mathsf{SU}(#1)}
\newcommand{\aisuff}[2]{\mathsf{SU}(#1,#2)}
\newcommand{\apref}[1]{\mathsf{PR}(#1)}
\newcommand{\arightquotient}[2]{\mathsf{RQ}(#1, #2)}
\newcommand{\afactors}[1]{\mathsf{FA}(#1)}

\def\defi{\mbox{\raisebox{0ex}[1ex][1ex]{$\stackrel{\mbox{\tiny def}}{\; =\;}$}}}

\newcommand{\ctobool}[1]{\code{toBool}(#1)}
\newcommand{\ctoint}[1]{\code{toInt}(#1)}
\newcommand{\ctostr}[1]{\code{toStr}(#1)}

\newcommand{\atostr}[1]{\mathsf{toStr}^\sharp(#1)}

\newcommand{\nodesize}{0.47}

\title{Static Program Analysis for String Manipulation Languages}
\author{Vincenzo Arceri
	\institute{University of Verona, Verona, Italy}
	\email{vincenzo.arceri@univr.it}
	\and
	Isabella Mastroeni
	\institute{University of Verona, Verona, Italy}
	\email{isabella.mastroeni@univr.it}
}
\def\titlerunning{Static Program Analysis for String Manipulation Languages}
\def\authorrunning{V. Arceri \& I. Mastroeni}

\begin{document}
\maketitle 

%% Abstract
\begin{abstract} In recent years, dynamic languages, such as JavaScript or Python, have been increasingly used in a wide range of fields and applications. Their tricky and misunderstood behaviors pose a hard challenge for static analysis of these programming languages. A key aspect of any dynamic language program is the multiple usage of strings, since they can be implicitly converted to another type value, transformed by string-to-code primitives or used to access an object-property. Unfortunately, string analyses for dynamic languages still lack precision and do not take into account some important string features. Moreover, string obfuscation is very popular in the context of dynamic language malicious code, for example, to hide code information inside strings and then to dynamically transform strings into executable code. In this scenario, more precise string analyses become a necessity. This paper is placed in the context of static string analysis by abstract interpretation and proposes a new semantics for string analysis, placing a first step for handling dynamic languages string features.
\end{abstract}

\section{Introduction}

Dynamic languages, such as JavaScript or Python, have faced an important increment of usage in a very wide range of fields and applications. Common features in dynamic languages are dynamic typing (typing occurs during program execution, at run-time) and implicit type conversion\cite{pradel2015}, lightening the development phase and allowing not to block the program execution in presence of unexpected or unpredictable situations.
Moreover, one important aspect of dynamic languages is the way strings may be used. In JavaScript, for example, strings can be either used to access property objects or transformed into executable code, by using the global function \code{eval}. In this way, dynamic languages provide multiple string features that simplify writing programs, allowing, at the same time, statically unpredictable executions which may make programs harder to understand \cite{pradel2015}. For this reason, string obfuscation (e.g., string splitting) is becoming one of the most common obfuscation techniques in JavaScript malware \cite{xu2012}, making it hard to statically analyze code. Consider, for example, the JavaScript program fragment in Fig.~\ref{fig:mal} where strings are manipulated, de-obfuscated, combined together into the variable \code{d} and finally transformed into executable code, the statement \code{ws = new ActiveXObject(WScript.Shell)}.
This command, in Internet Explorer, opens a shell which may execute malicious commands. The command is not hard-coded in the fragment but it is built at run-time and the initial values of $\code{i}$,$\code{j}$ and $\code{k}$ are unknown, such as the number of iterations of the loops in the fragment. These observations suggest us that, in order to statically understand statements dynamically generated and executed, it may be extremely useful to statically analyze the string value of \code{d}. Unfortunately, existing static analyzers for  dynamic languages \cite{tajs2009, jsai2014, safe2012,hauzar2015}, may fail to precisely analyze strings in dynamic contexts. For instance, in the example, existing static analyzers \cite{tajs2009, jsai2014, safe2012} lose precision on the \code{eval} input value, losing any information about it. Namely, the issue of analyzing dynamic languages, even if tackled by sophisticated tools as the cited ones, still lacks  
formal approaches for handling the more dynamic features of string manipulation, such as dynamic typing, implicit type conversion and dynamic code generation.

\noindent
\textbf{Contributions.} In this paper, we focus on the characterization of an abstract interpretation-based \cite{cousot1977} formal framework for handling dynamic typing and implicit type conversion, by defining an abstract semantics able to capture these dynamic features. Even if we do not tackle the problem of analyzing dynamically generated code (meaning that we do not analyze its \textit{behavior}), we strongly believe that such a semantics is a necessary step towards a sufficiently precise analysis of dynamically generated code, being able to reason about a class of string manipulation programs (as far as string values are concerned) that state-of-art static analyzers would fail to precisely analyze.
Indeed, the domain we propose allows us to \textit{collect} (and potentially approximate) the set of all the string values that a variable may receive during computation (at each program point). It should be clear that, in order to analyze {\em what} an {\tt eval} statement may execute, we surely need to (over-)approximate the set of precise values that its parameter may have.
Hence, we propose an approach aiming at defining a collecting semantics for strings. With this task in mind, we first discuss how to combine abstract domains of primitive types (strings, integers and booleans) in order to capture dynamic typing. Once we have such an abstract domain, we define on it an abstract semantics for a toy language, augmented with implicit type conversion, dynamic typing and some interesting string operations, whose concrete semantics is inspired by the JavaScript one. In particular, for each one of these operations we provide the algorithm computing its abstract semantics and we discuss their soundness and completeness.
%
%we propose a novel static string analysis based on finite state automata, an infinite abstract domain able to collect (and approximate) string values. This domain is not finite and computations on it may diverge; for this reason it can be equipped with a widening forcing and accelerating convergence \cite{silva2006}. We consider the $\mujs$ language, augmented with implicit type conversion, dynamic typing and  interesting string operations, whose concrete semantics is inspired by the JavaScript one.
%We consider a sub-language of JavaScript, $\mujs$, taking into account implicit type conversion, dynamic typing and  some basic string operations.
%We define the string-value static analysis in the abstract interpretation framework \cite{cousot1977}, by formalizing the abstract domain and by defining the abstract semantics of $\mujs$ on this domain. 
%Moreover, we provide the implementation of the finite state automata library and the implementation of the $\mujs$ static analyzer, showing the feasibility of the proposed approach. 
 
%  The main contributions are:
%\begin{itemize}
%	\item[$\bullet$] Formalization of the finite state automata abstract domain;
%	\item[$\bullet$] Definition of a sound abstract semantics for $\mujs$; 
%	\item[$\bullet$] Implementation of the finite state automata abstract domain library;
%	\item[$\bullet$] Implementation of the $\mujs$ static analyzer.
%\end{itemize}
%%

\begin{figure}
\begin{multicols}{2}
	\begin{lstlisting}[tabsize=2, basicstyle=\fontsize{9pt}{9pt}\selectfont\ttfamily]
vd, ac, la = "";
v = "wZsZ"; m = "AYcYtYiYvYeYXY";
tt = "AObyaSZjectB";
l = "WYSYcYrYiYpYtY.YSYhYeYlYlY"; 
	
while (i+=2 < v.length) 
	vd = vd + v.charAt(i);
	
while (j+=2 < m.length) 
	ac = ac + m.charAt(j);
	\end{lstlisting}
	\columnbreak
		\begin{lstlisting}[tabsize=2, basicstyle=\fontsize{9pt}{9pt}\selectfont\ttfamily]
ac += tt.substring(tt.indexOf("O"), 3);
ac += tt.substring(tt.indexOf("j"), 11);

while (k+=2 < l.length) 
	la = la + l.charAt(k);

d = vd + "=new " + ac + "(" + la + ")";
eval(d);
	\end{lstlisting}
\end{multicols}
\caption{A potentially malicious obfuscated JavaScript program.}
\label{fig:mal}
\end{figure}

\noindent
\textbf{Paper structure.} In Sect.~\ref{sect:background} we recall relevant notions on finite state automata and the core language we adopt for this paper and the finite state automata domain, highlighting some important operations and theoretical results, respectively. In Sect.~\ref{sect:abs-domain} we discuss and present two ways of combining abstract domains (for primitive types) suitable for dynamic languages. Then, In Sect.~\ref{sect:str-op-abs-sem} we present the novel abstract semantics for string manipulation programs. Finally, in Sect.~\ref{sect:rel}  we discuss the related work compared to this paper and we conclude the paper.

\section{Background}\label{sect:background}

\subsection{Basic notations and concepts}

\paragraph{String notation.} We denote by $\Sigma$ a finite alphabet of symbols, its Kleene-closure by $\Sigma^*$ and a string element by $\sigma \in \Sigma^*$. If $\sigma = \sigma_0\sigma_1\cdots\sigma_n$,
%\footnote{We use the string positions starting from $1$ instead of from $0$ in the sake of readability. Nevertheless, we are conscious that in JavaScript (and therefore also our implementation), string positions starts from $0$.}, 
the length of $\sigma$ is $|\sigma|=n+1$ and the element in the $i$-th position is $\sigma_i$. Given two strings $\sigma, \sigma' \in \Sigma^*$, $\sigma\sigma'$ is their concatenation. A language is a set of strings, i.e., $\lin \in \wp(\Sigma^*)$. We 
use the following notations: $\Sigma^i\defi\sset{\sigma\in\Sigma^*}{|\sigma|=i}$ and $\Sigma^{< i}\defi\bigcup_{j< i}\Sigma^j$. Given $\sigma\in\Sigma^*$, $i,j\in\mathbb{N}$ ($i\leq j\leq |\sigma|$) the substring between $i$ and $j$ of $\sigma$ is the string $\sigma_i\cdots\sigma_{j-1}$, and we denote it by $\substringf{\sigma}{i}{j}$. Let $\mathbb{Z}$ be the set of integers. We denote by $\Sigma^*_{\scriptsize{\cint}} \defi \{+,-, \epsilon\}\cdot\{0,1, \dots, 9\}^+$ the set of \textit{numeric strings}, i.e., strings corresponding to integers. $\mathcal{I}: \Sigma^*_{\scriptsize{\cint}} \rightarrow \mathbb{Z}$ maps numeric strings to the corresponding integers. Dually, we define the function $\mathcal{S}: \mathbb{Z} \rightarrow \Sigma^*_{\scriptsize{\cint}}$ that maps each integer to its numeric string representation (e.g., 1 is mapped to the string \code{"1"}, not \code{"+1"}, -5 is mapped to \code{"-5"}).

\noindent
\textbf{Regular languages and finite state automata.} We follow \cite{hopcroft1979} for automata notation. A finite state automaton (FA) is a tuple $\aut = (Q, q_0, \Sigma, \delta, F)$ where $Q$ is a finite set of states, $q_0 \in Q$ is the initial state, $\Sigma$ is a finite alphabet, $\delta \subseteq Q \times \Sigma \times Q$ is the transition relation and $F \subseteq Q$ is the set of final states. In particular, if $\delta:Q\times\Sigma\rightarrow Q$ is a function then $\aut$ is called deterministic FA (DFA).\footnote{We consider DFA also those FAs which are not complete, namely such that a transition for each pair $(q,a)$ ($q\in Q$, $a\in\Sigma$) does not exists. They can be easily transformed in a DFA by adding a sink state receiving all the missing transitions.} 
The class of languages recognized by FAs is the class of regular languages. We denote the set of all DFAs as \DFA.
%We denote with $\hat{\delta}: Q \times \Sigma^* \rightarrow \mathcal{P}(Q)$ the transitive closure of $\delta$. A string $\sigma \in \Sigma^*$ is accepted by an automaton $A$ if $\hat{\delta}(q_0, \sigma) \cap F \neq \emptyset$. 
Given an automaton $\aut$, we denote the language accepted by $\aut$ as $\lang(\aut)$. A language $\lin$ is  regular  iff there exists a FA $\aut$ such that $\lin = \lang(\aut)$. From the Myhill-Nerode theorem\cite{davis1994}, for each regular language there uniquely exists a minimum automaton, i.e., with the minimum number of states, recognizing the language. 
Given a regular language $\lin$, we denote by $\minimize(\lin)$ the minimum DFA $\aut$ s.t. $\lin=\lang(\aut)$.

\noindent
\textbf{The programming language.} We consider an $\mujs$ language (Fig.~\ref{fig:mujs-syntax}) that 
%The language $\mujs$ is an imperative core of JavaScript (Fig.~\ref{fig:mujs-syntax}). In particular, it 
contains representative string operations taken from the set of methods offered by the JavaScript built-in class \code{String}\cite{w3school-string}. Other JavaScript string operations can be modeled by composition of the given string operations or as particular cases of them. Primitive values are $\cval=\cstr\cup\cint\cup\cbool\cup\{\cnan\}$ with $\cstr\defi\Sigma^*$ (strings on the alphabet $\Sigma$), %$\cint\defi\nint$, 
$\cbool\defi\{\true,\false\}$ and $\cnan$ a special value denoting not-a-number.

\noindent
\textbf{Implicit type conversion.} In order to capture the semantics of the language $\mujs$, inspired by the JavaScript semantics, we need to deal with \textit{implicit type conversion}\cite{arceri2017}. For each primitive value, we define an auxiliary function converting primitive values to other primitive values (Fig.~\ref{fig:type-juggling}). Note that all the functions behave like the identity when applied to values not needing conversion, e.g., $\code{toInt}$ on integers. Then, $\code{toStr}: \cval \rightarrow \cstr$ maps any input value to its string representation; $\code{toInt}: \cval \rightarrow \cint \cup \{\cnan\}$ returns the integer corresponding to a value, when it is possible: For $\true$ and $\false$ it returns respectively $1$ and $0$, for strings in $\Sigma^*_{\footnotesize{\nint}}$ it returns the corresponding integer, while all the other values are converted to $\cnan$. 
%$1$ for $\true$, $0$ for $\false$ and the identity on $\cnan$ and integers, while if the input is a string then if it is a numeric string the function returns the corresponding integer, otherwise it returns $\cnan$. 
For instance, $\code{toInt}(``42") = 42$, $\code{toInt}(``42hello") = \cnan$.  Finally,
$\code{toBool}: \cval \rightarrow \cbool$ returns $\false$ when the input is $0$,
and $\true$ for all the other non boolean primitive values. For example, implicit type conversion is applied when the guards of \code{while} and \code{if} statements do not evaluate to booleans (e.g., \code{while (1)} \{x=x+1;\}, the guard is implicitly converted to \code{true}).

\begin{figure}[t]
	\begin{framed}
		\vbox{%
			\setlength{\grammarparsep}{3pt plus 1pt minus 1pt} % increase separation between rules
			\setlength{\grammarindent}{4em} % increase separation between LHS/RHS 
			\renewcommand{\syntleft}{}  \renewcommand{\syntright}{}
			{\footnotesize
			\begin{grammar}			<\expr> ::= \id ~|~ v $\in\cval$
				~|~ \expr  \ \mbox{\tt +} \expr
				~|~ \expr  \ \mbox{\tt -} \expr
				~|~ \expr  \ \mbox{\tt *} \expr
				~|~ \expr  \ \mbox{\tt /} \expr
				~|~ \expr \ \mbox{\tt \&\&} \expr
				\alt \expr \ \mbox{\tt ||} \expr
				~|~ \mbox{\tt !} \expr
				~|~ \expr  \ \mbox{\tt >} \expr
				~|~ \expr  \ \mbox{\tt \textless} \expr
				~|~ \expr \ \mbox{\tt ==} \expr
				~|~ \expr\lstinline{.}\substring{\expr}{\expr}
				\alt \expr\lstinline{.}\charat{\expr}
				~|~ \expr\lstinline{.}\indexof{\expr}	
				~|~ \expr\lstinline{.}\length
				
				<\block> ::= \lstinline@{@ \lstinline@}@
				~|~ \lstinline@{@ \stmt \ \lstinline@}@
				
				<\stmt> ::=	  \id \ \mbox{\tt =} \expr \mbox{\tt ;}
				~|~ \mbox{\tt if} (\expr) \block \ \mbox{\tt else} \block
				~|~ \mbox{\tt while} (\expr) \block	
				~|~ \block	
				~|~ \stmt \ \stmt
				~|~ \ulitright{;}
				\vspace{-0.5cm}
			\end{grammar}}
		}%
	\end{framed}
	\caption{$\mujs$ syntax}
	\label{fig:mujs-syntax}
\end{figure}

\noindent
\textbf{Semantics.} Program states are partial maps from identifiers to primitive values, i.e., $\states: \id \rightarrow \cval$. The concrete big-step semantics $\sem{\cdot} : \stmt \times \states \rightarrow \states$ follows~\cite{arceri2017}, and it includes dynamic typing and implicit type conversion. Also the expression semantics, $\sem{\cdot}:\expr\times\states\rightarrow\cval$, is standard; we only provide the formal and precise semantics of the $\mujs$ string operations. Let $\sigma, \sigma'\in\cstr$ and $i,j\in\cint$ (values which are not strings or numbers respectively, are converted by the implicit type conversion primitives. Negative values are treated as zero).
\begin{description}
	\item[{\tt substring}:] It extracts substrings from strings, i.e., all the characters between two indexes. The semantics is the function {\sc Ss}$: \cstr \times \cint \times \cint \rightarrow \cstr$  defined as: 
	\[
		\substringl{\sigma}{i}{j} \defi
		\begin{cases}
%			\semt{$\sigma.\substring{i}{0}$} & \mbox{if}\ j < 0 \\			
%			\semt{$\sigma.\substring{0}{j}$} & \mbox{if}\ i < 0 \\			
%			\semt{$\sigma.\substring{j}{i}$} & \mbox{if}\ j < i \\
			\substringl{\sigma}{j}{i}		&  j < i \\
			%\substringf{\sigma}{i}{j} &  j < |\sigma|\ \wedge\ i\leq j \\
			%\substringf{\sigma}{i}{n} &  j\geq n = |\sigma| \wedge\ i\leq j 
			\substringf{\sigma}{i}{\mathsf{max}(j,|\sigma|)} &  \mbox{otherwise}
		\end{cases}
	\]

	\item[{\tt charAt}:] It returns the character at a specified index. The semantics is the function {\sc Ca}$: \cstr \times \cint \rightarrow \cstr$ defined as follows:
	\[
	\charats{\sigma}{i} \defi \begin{cases}
	\sigma_i & 0 \leq i < |\sigma| \\
	\epsilon & \mbox{otherwise}
	\end{cases}
	\]
	\item[{\tt indexOf}:] It returns the position of the first occurrence of a given substring. The semantics is the function {\sc Io}$:\cstr \times \cstr \rightarrow \cint$ defined as follows: 
	\[
		\indexofs{\sigma}{\sigma'}\defi \begin{cases}
		\mathsf{min}\sset{i}{\sigma_i\dots\sigma_j = \sigma'} & \exists i,j.\:\sigma_i\dots\sigma_j = \sigma' \\
		-1 & \mbox{otherwise}
		\end{cases}
	\]
	\item[{\tt length}:] It returns the length of a string $\sigma \in \cstr$. Its semantics is the function {\sc Le}$:\cstr\rightarrow\cint$ defined as $\lengths{\sigma} \defi |\sigma| $.
	\item[{\tt concat}:] The string concatenation is handled by $\mujs$ plus operator (\code{+}). The concrete semantics relies on the concatenation operator reported in Sect.~\ref{sect:background}, i.e., $\concs{\sigma}{\sigma'} = \sigma\sigma'$.

\end{description}

\begin{figure}[t]
	\begin{minipage}{0.27\textwidth}
		\begin{adjustbox}{scale=0.76}
			$
			\ctostr{v} = 
			\begin{cases}
			v & v \in \cstr \\
			``\code{NaN}" & v = \cnan \\
			``\true" &  v = \true \\
			``\false" & v = \false \\ 
			\mathcal{S}(v) &  v \in \cint
			\end{cases}
			$
		\end{adjustbox}
	\end{minipage}
	~
	\begin{minipage}{0.27\textwidth}
		\begin{adjustbox}{scale=0.76}
			$
			\ctoint{v} =
			\begin{cases}
			v &  v \in \cint \\
			1 &  v = \true \\
			0 &  v = \false  \vee v = \cnan \\
			\mathcal{I}(v) &  v \in \cstr \wedge  v \in \Sigma^*_{\scriptsize{\cint}} \\ 
			\cnan & v \in \cstr \wedge v \not\in \Sigma^*_{\scriptsize{\cint}} 
			\end{cases}
			$
		\end{adjustbox}
	\end{minipage}
	~\hspace*{0.55cm}
	\begin{minipage}{0.25\textwidth}
		\begin{adjustbox}{scale=0.76}
			$
			\ctobool{v} =
			\begin{cases}
			v & v \in \cbool \\
			\true & v \in \cint \smallsetminus \{0\} \vee v \in \cstr \smallsetminus \{\epsilon\} \\
			\false & v = 0 \vee v = \epsilon \vee v = \cnan
			\end{cases}
			$
		\end{adjustbox}
	\end{minipage}
	\caption{$\mujs$ implicit type conversion functions.}
	\label{fig:type-juggling}
\end{figure}

\subsection{The finite state automata domain for strings}\label{sect:fa-domain}
In this section, we describe the automata abstract domain for strings \cite{park2016, wid-approach, yu2008}, namely the domain of regular languages over $\Sigma^*$. In particular, our aim is that of characterize automata as a domain for abstracting the computation of program semantics in the abstract interpretation framework.
The exploited idea is that of approximating strings as regular languages represented by the minimum DFAs~\cite{davis1994} recognizing them. In general, we have more DFAs that recognize a regular language, hence the domain of automata is indeed the quotient $\fa$ w.r.t. the equivalence relation induced by language equality: $\forall \aut_1,\aut_2\in\DFA.\:\aut_1 \equiv \aut_2 \Leftrightarrow \lang(\aut_1) = \lang(\aut_2)$. Hence, any equivalence class $[\aut]_{\equiv}$ is composed by the automata that recognize the same regular language.
We abuse notation by representing equivalence classes in the domain $\fa$ w.r.t.\ $\equiv$ by one of its automata (usually the minimum), i.e., when we write $\aut\in\fa$ we mean $[\aut]_{\equiv}$. The partial order $\leqfa$ induced by language inclusion is 
$\forall \aut_1, \aut_2 \in \fa \ . \ \aut_1 \leqfa \aut_2 \Leftrightarrow \lang(\aut_1) \subseteq \lang(\aut_2)$, which is well defined since automata in the same $\equiv$-equivalence class recognize the same language.
\noindent
The least upper bound (lub) $\lubfa: \fa \times \fa \rightarrow \fa$ on the domain $\fa$, corresponds to the standard union between automata:
$\forall \aut_1, \aut_2 \in  \fa.\:\aut_1 \lubfa \aut_2 \defi \minimize(\lang(\aut_1) \cup \lang(\aut_2))$.
It is the minimum automaton recognizing the union of the languages $\lang(\aut_1)$ and $\lang(\aut_2)$. This is a well-defined notion since regular languages are closed under union.
%As example, consider Fig.~\ref{fig:fa-lub}, where the automaton in Fig.~\ref{fig:fa-min} is the lub of $\aut_1$ and $\aut_2$ given in Fig.~\ref{fig:1-fa} and Fig.~\ref{fig:2-fa}, respectively. 
The greatest lower bound $\glbfa : \fa \times \fa \rightarrow \fa$ corresponds to automata intersection, since regular languages are closed under finite intersection:
$\forall \aut_1, \aut_2 \in  \fa.\: \aut_1 \glbfa \aut_2 \defi \minimize(\lang(\aut_1) \cap \lang(\aut_2)).$
%
%Also the greatest lower bound is well-defined since regular languages are closed under intersection.
%
%\begin{lemma} $\latticefa$ is a lattice.\end{lemma}
%\begin{proof}
%The lemma holds as regular languages (hence, finite state automata) are closed under union and intersection, i.e. $\lubfa$ and $\glbfa$.
%\end{proof}
%
%
%\vspace{-.75cm}
\begin{theorem}\label{thm:fa-moore-family}$\latticefa$ is a sub-lattice but not a complete meet-sub-semilattice of $\wp(\Sigma^*)$.\end{theorem} 
In other words, there exists no Galois connections between $\fa$ and $\wp(\Sigma^*)$, i.e., there may exist no minimal automaton abstracting a language.\footnote{Note that, some works have studied automatic procedures to compute, given an input language $L$, the \textit{regular cover} of $L$ \cite{domaratzki2001} (i.e., an automaton containing the language $L$). In particular, \cite{campeanu2002, domaratzki2001} have studied regular covers guaranteeing that the automaton obtained is the best w.r.t. a \textit{minimal relation} (but not minimum).}
However, this is not a concern, since the relation between concrete semantics and abstract semantics can be weakened while still ensuring soundness \cite{cousot1992}. A well known example is the convex polyhedra domain \cite{cousot1978}.

\noindent
\textbf{Widening.} The domain $\fa$ is an infinite domain, and it is not ACC, i.e., it contains infinite ascending chains. For instance, consider the set of languages $\{\sset{a^j b^j}{0\leq j\leq i}\}_{i\geq 0}\subseteq\wp(\Sigma^*)$ forming an infinite ascending chain, then also the set of the corresponding minimal automata forms an ascending chain on $\fa$. This clearly implies that any computation on $\fa$ may lose convergence \cite{cousot1992}.
Most of the proposed abstract domains for strings \cite{costantini2015,jsai2014, tajs2009, safe2012} trivially satisfy ACC by being finite, but they may lose precision during the abstract computation \cite{cousot1992-2}. 
%It is well known \cite{cousot1992-2} that infinite domains equipped with widening/narrowing are usually more precise. 
In these cases, domains must be equipped with a widening operator approximating the lub in order to force convergence (by necessarily losing precision) for any increasing chain \cite{cousot1992-2}.
As far as automata are concerned, existing widenings are defined in terms of a state equivalence relation merging states that recognize the same language, up to a fixed length $n$ (set as parameter for tuning the widening precision) \cite{silva2006,DBLP:conf/cav/BartzisB04}. We denote  this parametric widening with $\nabla_n$, $n \in \mathbb{N}$\cite{silva2006}.

\begin{example} Consider the following $\mujs$ fragment
	
	\begin{CenteredBox}
		\begin{lstlisting}[numbers=none]
		str = ""; while (x++ < 100) { str += "a"; }
		\end{lstlisting}
	\end{CenteredBox}
	
Since the value of the variable \code{x} is unknown, also the number of iterations of the \code{while}-loop is unknown. In these cases, in order to guarantee soundness and termination, we apply the widening operator. In Fig.~\ref{fig:1-wid} we report the abstract value of the variable \code{str} at the beginning of the second iteration of the loop, while in Fig.~\ref{fig:2-wid} the abstract value of the variable \code{str} at the end of the second iteration is reported. Before starting a new iteration, in the example, we apply $\nabla_1$ between two automata, namely we merge all the states having the same outgoing character. The minimization of the obtained automaton is reported in Fig.~\ref{fig:3-wid}. The next iteration will reach the fix-point, guaranteeing  soundness and termination.
	
\begin{figure}[t]
	\centering
		(a)\begin{subfigure}[t]{0.25\textwidth}
		\centering
		\begin{adjustbox}{scale=0.85} 
			\begin{tikzpicture}[->,>=stealth',shorten >=1pt,auto,node distance=2cm, semithick]
			\node[initial,state,scale=0.45, accepting, initial text =] 			 	 (A)                    {};
			\node[state,scale=0.45, accepting]        			 				     (B) [right of=A] 		{};
				
			\path[->] (A) edge node {$a$} (B);
			\end{tikzpicture}
		\end{adjustbox}
	\caption{}
	\label{fig:1-wid}
\end{subfigure}%
		~ 
(b)\begin{subfigure}[t]{0.3\textwidth}
	\centering
	\begin{adjustbox}{scale=0.85} 
		\begin{tikzpicture}[->,>=stealth',shorten >=1pt,auto,node distance=2cm, semithick]
			\node[initial,state,scale=0.45, initial text =] 			 	 (A)                    {};
			\node[state,scale=0.45, accepting]      			 					 (B) [right of=A] 		{};
			\node[state,scale=0.45, accepting]        			 				     (C) [right of=B] 		{};
				
			\path[->] (A) edge node {$a$} (B);
			\path[->] (B) edge node {$a$} (C);
		\end{tikzpicture}
	\end{adjustbox}
	\caption{}
	\label{fig:2-wid}
\end{subfigure}
~
(c)\begin{subfigure}[t]{0.3\textwidth}
	\centering
	\begin{adjustbox}{scale=0.85} 
		\begin{tikzpicture}[->,>=stealth',shorten >=1pt,auto,node distance=2cm, semithick]
			\node[initial,state,scale=0.45, initial text =, accepting] 				 (A)                   {};
				
			\path[->] (A) edge [loop above] node {$a$} (A);
		\end{tikzpicture}
	\end{adjustbox}
	\caption{}
	\label{fig:3-wid}
\end{subfigure}
\caption{(a) $\aut_1$ s.t. $\lang(\aut_1) = \{\epsilon, a\}$ (b)$\aut_2$ s.t. $\lang(\aut_2) = \{a, aa\}$ (c) $\aut_1 \nabla_1 \aut_2$}
\label{fig:wid}
\end{figure}
\end{example}

\section{An abstract domain for string manipulation}\label{sect:abs-domain}
In this section, we discuss how to design an abstract domain for string manipulation dealing also with other primitive types, namely able to combine different abstractions of different primitive types.
In particular, since operations on strings combine strings also with other values (e.g., integers), an abstract domain for string analysis equipped with dynamic typing must include all the possible primitive values, i.e., the whole $\cval=\cint\cup\cbool\cup\cstr\cup\{\cnan\}$.
The idea is to consider an abstract domain for each type of primitive value and to combine these abstract domains in a unique abstract domain for $\cval$. 
Consider, for each primitive value $\mathbb{D}$, an abstract domain $\mathbb{D}^\sharp$ (we denote the domain $\mathbb{D}^\sharp$ without bottom as $\mathbb{D}^\sharp_{\not\bot}$), equipped with an abstraction $\alpha_{\mathbb{D}}: \mathbb{D} \rightarrow \mathbb{D}^\sharp$ and a concretization $\gamma_{\mathbb{D}}:\mathbb{D}^\sharp \rightarrow \mathbb{D}$ forming a Galois insertion \cite{cousot1977}.

\noindent
\textbf{Coalesced sum.} One way to merge domains is the \textit{coalesced sum} \cite{cousot1997}. The resulting domain contains all the non-bottom elements of the domains, together with a new top and a new bottom, covering all the elements and covered by all the elements, respectively. In our case, if we consider the abstract domains $\cint^\sharp$, $\cstr^\sharp$ and $\cbool^\sharp$, the coalesced sum is the abstraction of $\wp(\cval)$ depicted in Fig.~\ref{fig:sum-domain}.
\begin{figure}[t]
	\centering
	\begin{tikzpicture}[scale=0.55]
	\node (top) at (0,1) {$\top$};
	\node (int) at (2,0) {$\astr_{\not\bot}$};
	\node (bool) at (-2,0) {$\abool_{\not\bot}$};
	\node (string) at (5,0) {$\{\anan\}$};
	\node (null) at (-5,0) {$\aint_{\not\bot}$};
	\node (bot) at (0,-1) {$\bot$};
	\draw (bot) -- (int) -- (top) -- (bool) -- (bot);
	\draw (top) -- (string) -- (bot) -- (null) -- (top);
	\end{tikzpicture}
	\caption{Coalesced sum abstract domain for $\mujs$}
	\label{fig:sum-domain}
\end{figure}
%
%\begin{wrapfigure}{l}{0.3\textwidth} 
%	\vspace{-.6cm}
%	\begin{CenteredBox}
%		\begin{lstlisting}
%		if (y < 5)
%			x = "42";
%		else
%			x = true;
%		\end{lstlisting}
%	\end{CenteredBox}
%	\vspace{-.5cm}
%\end{wrapfigure}
This is the simplest choice, but unfortunately this is not suitable for dynamic languages, and in particular for dealing with dynamic typing and implicit type conversion. The problem is that the type of variables is inferred at run-time and may change during execution. For example, consider the following $\mujs$ fragment: 
$\mathtt{if\ (y < 5)\ \{x = ``42";\}\ else\ \{x = true;\}}$.
The value of the variable $\code{y}$ is statically unknown hence, in order to guarantee soundness, we must take into account both the branches, meaning that \code{x} may be both a string and a boolean value, after the \code{if} statement. On the coalesced sum domain, the analysis would lose any precision w.r.t. collecting semantics by returning $\alpha_{\scriptsize{\cstr}}(``\code{42}") \sqcup \alpha_{\scriptsize{\cbool}}(\code{true}) = \top$. 

\noindent
\textbf{Cartesian product.}
In order to catch union types, without losing too much precision, we need to {\em complete} \cite{GRS00,GQ01,GM16} the above domain in order to observe collections of values of different types. 
In order to define this combination, we rely on the cartesian product, following \cite{fromherz18}.
%, i.e., given $X\subseteq\wp(\mathbb{X})$ and $Y\subseteq\wp(\mathbb{Y})$ ($\mathbb{X}$ and $\mathbb{Y}$ arbitrary sets), we denote it as $X \times Y\defi\sset{(x,y)}{x\in X,\ y\in Y}\subseteq\wp(\mathbb{X}\times\mathbb{Y})$. 
Hence, the complete abstract domain  w.r.t. dynamic typing and implicit type conversion is:
$\aint \times \abool \times \astr \times \wp(\{\anan\})$, % \in\uco(\wp(\cval))
abstraction of $\wp(\cval)$.
In this combining abstract domain, the value of \code{x} after the \code{if}-execution is precisely $(\bot, \alpha_{\scriptsize{\cbool}}(\code{true}), \alpha_{\scriptsize{\cstr}}(``\code{42}"), \bot)$, now an element of the domain, inferring that the value of \code{x} can be $\alpha_{\scriptsize{\cbool}}(\code{true})$ or $\alpha_{\scriptsize{\cstr}}(``\code{42}")$, but definitely not an abstract integer~or~$\cnan$.

In the following, we consider the abstract domain $\aval$ for string analysis obtained as cartesian product of the following abstractions:
%in sake of simplicity, we will use the coalesced sum domain for the soundness and completeness proofs and to focus the attention on the abstract semantics of string operations. However, the abstract semantics provided in the next sections can be lifted in the reduced product abstract domain. The instantiation of the coalesced sum domain of $\mujs$, for each primitive type, is the following.
%
%\[
%\begin{array}{lll} 
%\aint = \interval & \qquad\astr = \fa  &\qquad
% \abool = \wp(\{\code{true}, \code{false}\})%& \qquad \anan = \wp(\code{\{NaN\}})
%\end{array}
%\]
%
$\aint = \interval$ (the well-known abstract domain of intervals \cite{cousot1977}), $\astr = \fa$, $\abool = \wp(\{\code{true}, \code{false}\})$.  
%In the next section we define the abstract semantics of string operation of $\mujs$. As shown in Section \ref{sect:fa-domain}, $\fa$ does not form a Galois connection with $\mathcal{P}(\Sigma^*)$. Anyway, we will show that it can be fitted in the abstract interpretation framework, proving the soundness of the string operations abstract semantics and their closure on $\fa$.

\section{The $\mujs$ abstract semantics}\label{sect:str-op-abs-sem}

In this section, we define the abstract semantics of the language $\mujs$ on the abstract domain $\aval$. 
%It is well known that the semantics of statements does not depend on the considered abstract domain, while we need to define the abstract semantics of expressions since they produce values, and therefore abstract values. 
In particular, we have to define the expressions abstract semantics $\asem{\cdot}:\expr \times \states \rightarrow \aval$, which is standard except for  the string operations that will be explicitly provided by describing the algorithm for computing them. 
%For \code{substring} and \code{charAt}, we first define a \textit{concrete} operator on finite state automata, without abstractions on the other primitive types. These operator will be used in defining their abstract semantics. Then, for each operator, we discuss about soundness and completeness. We abuse notation defining the abstract semantics of expression as $\asem{\cdot}\cdot:\expr \times \states \rightarrow \cval$.
Let us first recall some important notions on regular languages, useful for the algorithms we will provide.
\begin{definition}[Suffixes and prefixes\cite{davis1994}]\label{def:su-pr} Let $\lin\in\wp(\Sigma^*)$ be a regular language. The suffixes of $\lin$ are 
	$\mbox{\sc Su}(\lin) \defi \sset{y\in\Sigma^*}{\exists x\in\Sigma^*.\:xy \in \lin}$, 
	and the prefixes of $\lin$ are 
	$\mbox{\sc Pr}(\lin) \defi \sset{x\in\Sigma^*}{\exists y\in\Sigma^*.\:xy \in \lin}$.
\end{definition}
We can define the suffixes from a position, namely given $i \in \mathbb{N}$, the set of suffixes from $i$ is $\isuff{\lin}{i}$ $\defi$ $\sset{y\in\Sigma^*}{\exists x\in\Sigma^*.\:xy \in \lin, |x| = i}$. For instance, let $\lin = \{abc, hello\}$, then $\isuff{\lin}{2} = \{c, llo\}$. 

%\begin{definition}[Left quotient\cite{davis1994}]\label{def:left-quotient} Let $\lin_1,\lin_2\in\wp(\Sigma^*)$ be regular languages. The left quotient of $\lin_1$ w.r.t\ $\lin_2$ is
%	%
%	$\leftquotient{\lin_1}{\lin_2} \defi \sset{y\in\Sigma^*}{\exists x \in \lin_2.\:xy \in \lin_1}$.	
%	%
%\end{definition}
%
\begin{definition}[Right quotient\cite{davis1994}]\label{def:right-quotient} Let $\lin_1,\lin_2\in\Sigma^*$ be regular languages. The right quotient of $\lin_1$ w.r.t. $\lin_2$ is 
	$\rightquotient{\lin_1}{\lin_2} \defi \sset{x\in\Sigma^*}{\exists y \in \lin_2.\:xy \in \lin_1}$.
\end{definition}

\noindent
%For example, let $\lin_1 = \{xab, yac\}$ and $\lin_2 = \{x, y\}$. The left quotient of $\lin_1$ w.r.t\ $\lin_2$ is $\leftquotient{\lin_1}{\lin_2} = \{ ab, ac \}$. 
For example, let $\lin_1 = \{xab, yab\}$ and $\lin_2 = \{b, ab\}$. The right quotient of $\lin_1$ w.r.t.\ $\lin_2$ is $\rightquotient{\lin_1}{\lin_2} = \{ xa, ya, x, y \}$. 
%We now introduce a characterization of a subset of suffixes that we will use in the following.
%
%\begin{proposition} Let $\lin\in\wp(\Sigma^*)$ be a regular language. Given $i \in \mathbb{N}$, the set of suffixes from $i$ is defined as follows:
%	%
%	\[
%	\isuff{\lin}{i} \defi \sset{y\in\Sigma^*}{\exists x\in\Sigma^*.\:xy \in \lin, |x| = i} = \leftquotient{\lin}{\pref{\lin} \cap \Sigma^i}
%	\]
%	%
%\end{proposition}

\begin{definition}[Factors\cite{bordihn09}] Let $\lin\in\wp(\Sigma^*)$ be a regular language. The set of its factors is
	$	\factors{\lin} \defi \sset{y\in\Sigma^*}{\exists x,z \in \Sigma^*.\:xyz \in \lin}	$.
\end{definition}
%\begin{proposition} Let $\lin\in\wp(\Sigma^*)$. The following fact trivially holds
%	%
%	\[
%	\factors{\lin} = \suff{\pref{\lin}} = \pref{\suff{\lin}}
%	\]
%	%
%\end{proposition}
These operations are all defined as transformations of regular languages. In \cite{davis1994} the corresponding algorithms on FA are provided. In particular, let $\aut,\aut_1\in\fa$ and $i \in \mathbb{N}$, then $\asuff{\aut}$, $\apref{\aut}$, $\aisuff{\aut}{i}$, $\afactors{\aut}$ and $\arightquotient{\aut}{\aut_1}$  
%$\aleftquotient{\aut}{\aut_1}$
 are the algorithms corresponding  to the transformations $\suff{\lang(\aut)}$, $\pref{\lang(\aut)}$, $\isuff{\lang(\aut)}{i}$, $\factors{\lang(\aut)}$ and  $\rightquotient{\lang(\aut)}{\lang(\aut_1)}$, respectively. 
%and $\leftquotient{\lang(\aut)}{\lang(\aut_1)}$, respectively. 
%Moreover, in \cite{davis1994} the authors proved that regular languages are closed under left-right quotient, suffix and prefix operations, as well as under the suffixes from $i \in \mathbb{N}$.
%
%Later, we will refer to such operation on automata, given $A,B \in \fa, i \in \mathbb{Z}$, with $\asuff{A}$, $\apref{A}$, $\aisuff{A}{i}$, $\afactors{A}$, $\arightquotient{A}{B}$, $\aleftquotient{A}{B}$.
%It has been proved that regular languages are closed under left-right quotient, suffix and prefix operations\cite{davis1994}, as well as the suffixes up to $i \in \mathbb{N}$, as it defined as composition of closed operation. 
%
%Moreover, regular languages are closed under all the above reported operation, that is, 
Namely, $\forall \aut,\aut_1\in\fa$, $i \in \mathbb{N}$, the following facts hold:
\hspace{-.5cm}
{\small\[
	\begin{array}{l}
	\suff{\lang(\aut)} = \lang (\asuff{\aut}),\ \pref{\lang(\aut)} = \lang (\apref{\aut}),\ \factors{\lang(\aut)} = \lang(\afactors{\aut})\\
	\rightquotient{\lang(\aut)}{\lang(\aut_1)} = \lang(\arightquotient{\aut}{\aut_1}),\  \isuff{\lang(\aut)}{i} = \lang(\aisuff{\aut}{i})%, \\
%	\leftquotient{\lang(\aut)}{\lang(\aut_1)} = \lang(\aleftquotient{\aut}{\aut_1})
	\end{array}
	\]}
%{\small\[
%\begin{array}{lr}
%\suff{\lang(\aut)} = \lang (\asuff{\aut}) & \isuff{\lang(\aut)}{i} = \lang(\aisuff{\aut}{i}) \\
%\pref{\lang(\aut)} = \lang (\apref{\aut}) &  \factors{\lang(\aut)} = \lang(\afactors{\aut})\\
%\rightquotient{\lang(\aut)}{\lang(\aut_1)} = \lang(\arightquotient{\aut}{\aut_1}) &  \\
%\leftquotient{\lang(\aut)}{\lang(\aut_1)} = \lang(\aleftquotient{\aut}{\aut_1})
%\end{array}
%\]}

As far as (state) complexity is concerned\cite{YuZS94}, prefix and right quotient operations have linear complexity, while suffix and factor operations, in general, are exponential\cite{YuZS94,pribavkina2010}.
\subsection{Abstract semantics of {\tt substring}} 

In this section, we define the abstract semantics of \code{substring}, i.e., we define the operator {\sf SS}$^\sharp : \fa \times \interval \times \interval \rightarrow \fa$,  starting from an automaton, an interval $[i,j]$ of initial indexes and an interval $[l,k]$ of final indexes for substrings, and computing the automaton recognizing the set of all substrings of the input automata language between the indexes in the two intervals.
Hence, since the abstract semantics has to take into account the swaps when the initial index is greater than the final one, several cases arise handling (potentially unbounded) intervals. 
Tab.~\ref{tab:subs-abstract} reports the abstract semantics of {\sf SS}$^\sharp$ when $i,j \leq l$ (hence $i \leq k$). 
The definition of this semantics is by recursion with four base cases (the other cases are recursive calls splitting and rewriting the input intervals in order to match or to get closer to base cases) for which we describe the algorithmic characterization.
Consider $\aut\in\fa$, $i,l\in\nint\cup\{-\infty\}$, $j,k\in\nint\cup\{+\infty\}$ (for the sake of readability we denote by $\sqcup$ the automata lub $\lubfa$, and by $\sqcap$ the glb $\glbfa$), the base cases are
\begin{enumerate*}
\item If $i,j,l,k \in\mathbb{Z}$ (first row, first column of Tab.~\ref{tab:subs-abstract}) we have to compute the language of all the substrings between an initial index in $[i,j]$ and a final index in $[l,k]$, namely $\substringl{\lang(\aut)}{[i,j]}{[l,k]}\footnote{We abuse notation by denoting with $ \mbox{\sc Ss}$ also the additive lift to languages and to sets of indexes: $\mbox{\sc Ss}:\wp(\Sigma^*) \times \wp(\mathbb{Z}) \times \wp(\mathbb{Z}) \rightarrow \wp(\Sigma^*)$ defined as $\substringl{\lin}{I}{J} = \sset{\substringl{\lin}{i}{j}}{i \in I, j \in J}=\sset{\substringl{\sigma}{i}{j}}{\sigma,\in\lin, i \in I, j \in J}$.}$.% \defi $ $\bigcup_{a\in[i,j],b\in[l,k]} \sset{\substringl{\sigma}{a}{b}}{\sigma \in \lin}$. 
For example, let  $\lin = \{a\}^* \cup \{hello, bc\}$, the set of its substrings from  1 to  3 is $\substringl{\lin}{[1,1]}{[3,3]}$ = $\{\epsilon, a, aa, el, c\}$. 
The automaton accepting this language is computed by the operator
\[
\substringa{\aut}{[i,j]}{[l,k]}\defi\bigsqcup_{a\in[i,j],b\in[l,k]}(\arightquotient{\aisuff{\aut}{a}}{\aisuff{\aut}{b}}\sqcap \minimize(\Sigma^{b-a}))\sqcup (\aisuff{\aut}{a}) \sqcap \minimize(\Sigma^{<b-a})
\]

\item When both intervals correspond to $[-\infty, +\infty]$, the result is the automaton of all possible factors of $\aut$ (last row, last column), i.e.,  $\afactors{\aut}$;
\item If $[i,j]$ is defined and the interval of final indexes is unbounded, i.e., $[l, +\infty]$ (first row, third column), we have to compute  the automaton recognizing the following language
\[
\substringlright{\lang(\aut)}{[i,j]}{l}\defi\bigcup_{a\in[i,j]}\sset{\substringl{\sigma}{a}{k}}{\sigma \in\lang(\aut),\ k \geq l}
\]
\noindent
i.e., all the strings between a finite interval of initial indexes and an unbounded final index.
The automaton accepting this language is computed by
\begin{equation*}
\substringaright{\aut}{[i,j]}{l}\defi\bigsqcup_{a\in[i,j]}\arightquotient{\aisuff{\aut}{a}}{\asuff{\aisuff{\aut}{l}}}
\end{equation*}
The abstract semantics returns the least upper bound of all the automata of substrings from $a$ in $[i,j]$ to an unbounded index greater than or equal to $l$;
\item When both intervals are unbounded ($[i, +\infty]$ and $[l, +\infty]$, third row, third column of Tab.~\ref{tab:subs-abstract}), we 
split the language to accept. In particular, we compute the substrings between $[i, l]$ and $[l +\infty]$ (and this has been considered in case 3), and the automaton recognizing the language of all substrings with both initial and final index with any value greater than $l$, i.e., the language $\substringlr{\lang(\aut)}{l}$ $\defi \sset{\substringl{\sigma}{a}{b}}{\sigma \in \lang(\aut),\ a,b \geq l}$. This latter set is computed by the algorithm 
$\substringalr{\aut}{l}\defi\afactors{\aisuff{\aut}{l}}$
\end{enumerate*}
Here we show the table only for the case $i,j\leq l \ (\mbox{and thus } i \leq k)$. Only few cases are not considered and they are not reported for space limitations. Anyway, they are compatible with Tab.~\ref{tab:subs-abstract}.
% reported in Tab.~\ref{tab:subs-abstract2} and Tab.~\ref{tab:subs-abstract3} in the appendix.
In Fig.~\ref{fig:subs} we report an example obtained applying the rules in the tables.

%
%For completeness of the formal semantics of substring,  respectively provide the case when $l < i \leq k$ and $(i > k) \vee (i \leq l, j \geq l)$. Even in these cases, the semantics split and rewrite the input intervals reducing each recursive call to an already defined case.
\begin{table}[t]
\begin{adjustbox}{scale=0.75}
		\bgroup
		\def\arraystretch{2}%  1 is the default, change whatever you need
		\hspace*{0.5cm}		
		\begin{tabular}{ | c || c | c | c | c |}
		\hline
		 \makecell{$\asubstring{\aut}{[i,j]}{[l,k]}$\\$i,j\leq l\ (i\leq k)$}& $l,k \in \mathbb{Z}$  & \makecell{$l = -\infty$, $k \in \mathbb{Z}$} & \makecell{$l \in \mathbb{Z}$, $k = + \infty$} & \makecell{$l = -\infty$, $k = + \infty$} \\ 
		\hline
		\hline
		$i,j \in \mathbb{Z}$ & %$\ds{\bigcup_{\substack{a \in[i,j], b \in[l,k]}}}\substringa{\aut}{a}{b}$  
		$\substringa{\aut}{[i,j]}{[l,k]}$
		& $\asubstring{\aut}{ [i,j]}{[0,k]}$ & %$\ds{\bigcup_{a\in[i,j]}}\substringaright{\aut}{a}{l}$ 
		$\substringaright{\aut}{[i,j]}{l}$
		& $\asubstring{\aut}{[i,j]}{[0,+\infty]}$ \\ 
		\hline
		\makecell{$i = -\infty$, $j \in \mathbb{Z}$} & $\asubstring{\aut}{[0,j]}{[l,k]}$ & $\asubstring{\aut}{[0,j]}{[0,k]}$ & $\asubstring{\aut}{ [0,j]}{[l,+\infty]}$ & $\asubstring{\aut}{[0,j]}{[0,+\infty]}$\\ 
		\hline
		\makecell{$i \in \mathbb{Z}$, $j = + \infty$} &  \makecell{
		%$\ds{\bigcup_{a\in[l,k]}\substringaright{\aut}{a}{k}}
		$\asubstring{\aut}{[l,k]}{[k,+\infty]}
		$\\$\:\sqcup\:\asubstring{\aut}{[i,k]}{[l,k]}$} 		
		& $\asubstring{\aut}{[i,+\infty]}{[0,k]}$ & 
%		$
%		\begin{cases}
%			\asuff{\apref{\aisuff{A}{i}}}	& i =l \\
%			\asubstring{A}{[i, l]}{[l, +\infty]}\cup \asubstring{A}{[l, +\infty]}{[l, +\infty]} & i < l \\
%			\asubstring{A}{[l, i]}{[i, +\infty]}\cup \asubstring{A}{[i, +\infty]}{[i, +\infty]} & i > l \\
%		\end{cases} 
%		$ 
\makecell{%$\ds{\bigcup_{a\in[i,l]}}\substringaright{\aut}{a}{l}
$\substringaright{\aut}{[i,l]}{l}%$\asubstring{\aut}{[i,l]}{[l,+\infty]}$
\:\sqcup\:\substringalr{\aut}{l}$}
		& $\asubstring{\aut}{[i,+\infty]}{[0,+\infty]}$ \\ 
		\hline
		\makecell{$i = -\infty$, $j = + \infty$} & $\asubstring{A}{[0,+\infty]}{[l,k]}$ & $\asubstring{\aut}{[0,+\infty]}{[0,k]}$ & $\asubstring{\aut}{[0,+\infty]}{[l,+\infty]}$ & $\afactors{\aut}$ \\
		\hline
		\end{tabular}
		\egroup
\end{adjustbox}
\caption{Definition of $\mathsf{SS}^\sharp$ when $i,j\leq l \ (\mbox{and thus } i \leq k)$}
\label{tab:subs-abstract}
\end{table}

\begin{theorem}[Termination of $\mathsf{SS}^\sharp$] For each $\aut \in \fa, I, J \in \interval.\:\asubstring{\aut}{I}{J}$ performs at most three recursive calls, before reaching a base case.
\end{theorem}

\begin{theorem}$\mathsf{SS}^\sharp$ is sound and complete: $\forall \aut \in \fa, I, J \in \interval.\:\substringl{\lang(\aut)}{I}{J} = \lang(\asubstring{\aut}{I}{J})$.
\end{theorem}

\begin{figure}[t]
	(a)\begin{subfigure}[t]{0.475\textwidth}
		\centering		
		\begin{adjustbox}{scale=0.75} 
			\begin{tikzpicture}[->,>=stealth',shorten >=1pt,auto,node distance=1.9cm, semithick]
			\node[initial,state,scale=\nodesize, initial text =] 			 	 (A)                    {};
			\node[state, scale=\nodesize]      			 					 (F) [below right of=A] {};
			\node[state, scale=\nodesize]      			 					 (G) [right of=F] 		{};			\node[state, scale=\nodesize]      			 					 (H) [right of=G] 		{};			\node[state, scale=\nodesize]      			 					 (I) [right of=H] 		{};
			\node[state, scale=\nodesize]      			 					 (B) [above right of=A] {};
			\node[state, scale=\nodesize]      			 					 (C) [right of=B] 		{};			\node[state, scale=\nodesize]      			 					 (D) [right of=C] 		{};			\node[state, scale=\nodesize, accepting]      			 					 (E) [above right of=I] 		{};

			\path[->] (A) edge node {$l$} (B);
			\path[->] (B) edge node {$a$} (C);
			\path[->] (C) edge node {$n$} (D);
			\path[->] (D) edge node {$g$} (E);
			
			\path[->] (A) edge node {$h$} (F);
			\path[->] (F) edge node {$e$} (G);
			\path[->] (G) edge node {$l$} (H);
			\path[->] (H) edge node {$l$} (I);
			\path[->] (I) edge node {$o$} (E);
			
			\end{tikzpicture}
		\end{adjustbox}
		\caption{}
		\label{fig:left-subs}
	\end{subfigure}
	(b)\begin{subfigure}[t]{0.475\textwidth}
		\centering
		\begin{adjustbox}{scale=0.75} 
			\begin{tikzpicture}[->,>=stealth',shorten >=1pt,auto,node distance=1.9cm, semithick]
					\node[initial,state,scale=\nodesize, initial text =] 			 	 (A)                    {};
		\node[state, scale=\nodesize]      			 					 (F) [below right of=A] {};
		\node[state, scale=\nodesize, accepting]      			 					 (G) [right of=F] 		{};			\node[state, scale=\nodesize, accepting]      			 					 (H) [right of=G] 		{};			\node[state, scale=\nodesize, accepting]      			 					 (I) [above right of=H] 		{};
		\node[state, scale=0.45]      			 					 (B) [above right of=A] {};
		\node[state, scale=0.45, accepting]      			 					 (C) [right of=B] 		{};

		\path[->] (A) edge node {$a$} (B);
		\path[->] (B) edge node {$n$} (C);
		\path[->] (C) edge node {$g$} (I);
		
		\path[->] (A) edge node {$e$} (F);
		\path[->] (F) edge node {$l$} (G);
		\path[->] (G) edge node {$l$} (H);
		\path[->] (H) edge node {$o$} (I);

			\end{tikzpicture}
		\end{adjustbox}
		\caption{}
		\label{fig:right-subs}
	\end{subfigure}
	\caption{(a) $\aut$, $\lang(\aut) = \{lang, hello\}$. (b) $\aut' = \asubstring{\aut}{[1,1]}{[3,+\infty]}$, $\lang(\aut') = \{an, ang, el, ell, ello\}$.}
	\label{fig:subs}
\end{figure}
\subsection{Abstract semantics of \code{charAt}}
%We first define the additive lift to languages and set of integers of \code{charAt}  with the function $\mbox{{\sc Ca}}:\wp(\Sigma^*) \times \wp(\mathbb{Z}) \rightarrow \wp(\Sigma^*)$.
%%
%$\charats{\lin}{I} \defi \sset{\charats{\lin}{i}}{\sigma \in \lin, i \in I}$
%%
The abstract semantics of \code{charAt} should return the automaton accepting the language of all the characters of strings accepted by an automaton $\aut$, in a position inside a given interval $[i,j]$: This  is computed by $\mathsf{CA}^\sharp: \fa \times \interval \rightarrow \fa$
\[
\charata{\aut}{[l,h]} \defi
\begin{cases}
\bigsqcup_{i \in [l,h]} \substringa{\aut}{[i,i]}{[i+1,i+1]} & l,h \in \mathbb{Z} \\
\charata{\aut}{[0, h]} \sqcup \minimize(\{\epsilon\}) & l = -\infty, h \in \mathbb{Z}, h \geq 0 \\
\minimize(\{\epsilon\}) & l = -\infty, h \in \mathbb{Z}, h < 0 \\
\minimize(\mathtt{chars}(\aisuff{\aut}{l})) \sqcup \minimize(\{\epsilon\}) & l \in \mathbb{Z}, l \geq 0, h = +\infty \\
\minimize(\mathtt{chars}(\aut)) \sqcup \minimize(\{\epsilon\}) & l = -\infty \mbox{ or } l \in \mathbb{Z}, l < 0, h = + \infty
\end{cases}
\]
We call $\mathsf{SS}$ (defined before) when the interval index $[l,h]$ is finite. 
%When $l <0$ or $l = -\infty$ and $h \in \mathbb{Z}$ is positive, we restrict the interval recursively calling $\mathsf{CA}^\sharp$ only on the positive values of the interval, adding $\minimize(\epsilon)$ as the result of the negative cases. A particular case is when $[l,h] \subseteq[-\infty, -1]$, where no valid index is inside the interval. In this case, we simply return $\minimize(\epsilon)$. 
In the last two cases, we use the function $\mathtt{chars}: \fa \rightarrow \wp(\Sigma)$, returning the set of characters read in any transition of an automaton. When $l \in \mathbb{Z}, h = +\infty$, we return the characters starting from $l$ together with $\minimize(\{\epsilon\})$ while, when $l=-\infty$, we simply return the characters of the automaton together with~$\minimize(\{\epsilon\})$.
\begin{theorem}\label{thm:charat} $\mathsf{CA}^\sharp$ is sound and complete: $\forall \aut \in \fa, I \in \interval $,
	%
	%\[
	$\charats{\lang(\aut)}{I} = \lang(\charata{\aut}{I})
	$%\]
	.\footnote{In the following, for all the string semantics, we abuse notation for the additive lift to languages and intervals.}
\end{theorem}
\subsection{Abstract semantics of \code{length}}
%We first consider the additive lift to languages of length operation. As we have done previously, we abuse notation by defining $\lengths{\lin} \defi \sset{\lengthf{\sigma}}{\sigma \in \lin}$. 
The abstract semantics of \code{length} should return the interval of all the possible string lengths in an automaton, i.e., it is  $\mathsf{LE}^\sharp : \fa \rightarrow \interval$ computed by Alg.~\ref{alg:length}, where $\mathsf{minPath}, \mathsf{maxPath}:\fa \times Q \times Q \rightarrow \wp(Q)$ return the minimum and the maximum paths between two states of the input automaton, respectively. $\mathsf{len}: \wp(Q) \rightarrow \mathbb{N}$ returns the size of a path, and $\mathsf{hasCycle}:\fa \rightarrow \{\true, \false\}$ checks whether the automaton contains cycles.

%\begin{center}
%\begin{figure}[th]
%	\scalebox{0.85}{%
%	\begin{algorithm}[H]
%		\KwData{Deterministic finite state automaton $\aut = (Q, \Sigma, \delta, q_0, F)$} 
%		\KwResult{$\lengtha{\aut}$}
%		$\mathsf{P\_len} \leftarrow 0$;	$\mathsf{p\_len} \leftarrow \infty $ 
%		
%		\eIf{$\mathsf{hasCycle}(\aut)$}{
%			\ForEach{$q_f \in F$} {
%				$\mathsf{p} \leftarrow \mathsf{minPath}(\aut, q_0, q_f)$\;		
%				
%				\lIf{$\mathsf{len}(\mathsf{p}) < \mathsf{p\_len}$}{
%					$\mathsf{p\_len} \leftarrow \mathsf{len}(\mathsf{p})$
%				}	
%			}		
%			\textbf{return} $[\mathsf{p\_len}, +\infty]$\;			
%		}
%		{
%			\ForEach{$q_f \in F$} {
%				$\mathsf{p} \leftarrow \mathsf{minPath}(\aut, q_0, q_f)$; $\mathsf{P} \leftarrow \mathsf{maxPath}(\aut, q_0, q_f)$\;
%				
%				\lIf{$\mathsf{len}(\mathsf{p}) < \mathsf{p\_len}$}{
%					$\mathsf{p\_len} \leftarrow \mathsf{len}(\mathsf{p})$
%				}	
%				
%				\lIf{$\mathsf{len}(\mathsf{P}) > \mathsf{P\_len}$}{
%					$\mathsf{P\_len} \leftarrow \mathsf{len}(\mathsf{P})$
%				}
%			}
%			\textbf{return} $[\mathsf{p\_len}, \mathsf{P\_len}]$\;			
%		}
%		
%		\caption{$\mathsf{LE}^\sharp: \fa \rightarrow \interval$ algorithm}
%		\label{alg:length}
%	\end{algorithm}
%}
%\end{figure}
%%\end{center}

%In order to understand how to compute, in a sound way, the interval of lengths of strings accepted by an automaton, 
%Consider the (minimum) automaton $\aut_1$ in Fig.\ref{fig:left-length}: 
\noindent
The idea is to compute the minimum and the maximum path reaching each final state in the automaton (in Fig.~\ref{fig:left-length}, we obtain $3$ and $5$). Then, we abstract the set of lengths obtained so far into intervals (in the example, $[3,5]$). Problems arise when the automaton contains cycles. In this case, we simply return the undefined interval starting from the minimum path, to a final state, to $+\infty$. For example, in the automaton in Fig.~\ref{fig:right-length}, the length interval is $[3, +\infty]$. 
%The procedure implementing $\mathsf{LE}^\sharp$ is reported in Alg.~\ref{alg:length}.
%
\begin{figure}[t]
		(a)\begin{subfigure}[t]{0.475\textwidth}
			\centering
	\begin{adjustbox}{scale=0.75} 
			\begin{tikzpicture}[->,>=stealth',shorten >=1pt,auto,node distance=1.9cm, semithick]
			\node[initial,state,scale=\nodesize, initial text =] 			 	 (A)                    {};
			\node[state, scale=\nodesize]      			 					 (B) [above right of=A] {};
			\node[state, scale=\nodesize]      			 					 (C) [right of=B] 		{};
			
			\node[state,scale=\nodesize]        			 				     (E) [below right of=A] {};
			\node[state,scale=\nodesize]        			 				     (F) [right of=E] 		{};
			\node[state,scale=\nodesize]        			 				     (G) [right of=F] 		{};
			\node[state,scale=\nodesize]        			 				     (H) [right of=G] 		{};
			\node[state,accepting,scale=\nodesize]         					 (D) [above of=H]		{};
			
			\path[->] (A) edge node {$a$} (B);
			\path[->] (B) edge node {$b$} (C);
			\path[->] (C) edge node {$c$} (D);
			
			\path[->] (A) edge node {$h$} (E);
			\path[->] (E) edge node {$e$} (F);
			\path[->] (F) edge node {$l$} (G);
			\path[->] (G) edge node {$l$} (H);
			\path[->] (H) edge node[swap] {$o$} (D);
			\end{tikzpicture}
		\end{adjustbox}
		\caption{}
		\label{fig:left-length}
	\end{subfigure}
	(b)\begin{subfigure}[t]{0.475\textwidth}
		\centering
		\begin{adjustbox}{scale=0.75} 
			\begin{tikzpicture}[->,>=stealth',shorten >=1pt,auto,node distance=1.9cm, semithick]
			\node[initial,state,scale=\nodesize, initial text =] 			 	 (A)                    {};
			\node[state, scale=\nodesize]      			 					 (B) [above right of=A] {};
			\node[state, scale=\nodesize]      			 					 (C) [right of=B] 		{};
			
			\node[state,scale=\nodesize]        			 				     (E) [below right of=A] {};
			\node[state,scale=\nodesize]        			 				     (F) [right of=E] 		{};
			\node[state,scale=\nodesize]        			 				     (G) [right of=F] 		{};
			\node[state,scale=\nodesize]        			 				     (H) [right of=G] 		{};
			\node[state,accepting,scale=\nodesize]         					 (D) [above of=H]		{};
			
			\path[->] (A) edge node {$a$} (B);
			\path[->] (B) edge node {$b$} (C);
			\path[->] (C) edge node {$c$} (D);
			
			\path[->] (A) edge node {$h$} (E);
			\path[->] (E) edge node {$e$} (F);
			\path[->] (F) edge node {$l$} (G);
			\path[->] (G) edge node {$l$} (H);
			\path[->] (H) edge node[swap] {$o$} (D);
			\path[->] (C) edge [bend right=100] node[swap] {$b$} (A);
			\end{tikzpicture}
		\end{adjustbox}
		\caption{}
		\label{fig:right-length}
	\end{subfigure}
	\caption{(a) $\aut_1$, $\lang(\aut_1) = \{abc, hello\}$. (b) $\aut_2$, $\lang(\aut_2) = \{abc, hello\} \cup \sset{(abb)^nc}{n > 0}$.}
\end{figure}
\begin{theorem}\label{thm:length} $\mathsf{LE}^\sharp$ is sound but not complete: $\forall \aut \in \fa$
	%
	%\[
	$\lengths{\lang(\aut)} \subset \lengtha{\aut}
	$. 
%	where $\lengths{\lin}$ is the lift of the concrete semantics to languages.
\end{theorem}
\subsection{Abstract semantics of \code{indexOf}}
\begin{figure}[t]
	(a)\begin{subfigure}{0.5\textwidth}
		\centering
		\begin{adjustbox}{scale=0.75}	
			\begin{tikzpicture}[->,>=stealth',shorten >=1pt,auto,node distance=2cm, semithick]
			\node[state, initial, initial text =,scale=\nodesize]				 (I)					{};	
			\node[state,scale=\nodesize] 									   	 (A) [right of=I] 		{};
			\node[state,scale=\nodesize]      			 					 (B) [right of=A] 		{};
			\node[state,scale=\nodesize,accepting]        			 		 (C) [right of=B] 		{};

			\node[state,scale=\nodesize] 										 (E) [below of =A]       {};
			\node[state,scale=\nodesize]      			 					 (F) [right of=E] 		{};
			
			\path[->] (I) edge node {$a$} (A);
			\path[->] (A) edge node {$b$} (B);
			\path[->] (B) edge node {$c$} (C);
			
			\path[->] (I) edge node[swap] {$d$} (E);
			\path[->] (E) edge node {$d$} (F);
			\path[->] (F) edge node[swap] {$d$} (C);
			
			\path[->] (I) edge [bend left=50] node {$b$} (B);
			
			\end{tikzpicture}
		\end{adjustbox}
		\caption{}
		\label{fig:left-io}
	\end{subfigure}
	(b)\begin{subfigure}[t]{0.5\textwidth}
			\centering	
			\vspace*{-0.8cm}\hspace*{-1.5cm}
			\begin{adjustbox}{scale=0.75} 
			\begin{tikzpicture}[->,>=stealth',shorten >=1pt,auto,node distance=1.9cm, semithick]
			\node[initial,state,scale=\nodesize, initial text =] 			 	 (A)                    {};
			\node[state, scale=\nodesize]      			 					 (B) [above right of=A] {};
			\node[state, scale=\nodesize]      			 					 (C) [right of=B] 		{};
		
			\node[state,scale=\nodesize]        			 				     (E) [below right of=A] {};
			\node[state,scale=\nodesize]        			 				     (F) [right of=E] 		{};
			\node[state,scale=\nodesize]        			 				     (G) [right of=F] 		{};
			
			\node[state,accepting,scale=0.45]         					 (D) [above of=G]		{};
			\path[->] (A) edge node {$b$} (B);
			\path[->] (B) edge node {$c$} (C);
			\path[->] (C) edge node {$d$} (D);
			
			\path[->] (A) edge node {$a$} (E);
			\path[->] (E) edge node {$a$} (F);
			\path[->] (F) edge node {$a$} (G);
			\path[->] (G) edge node[swap] {$b$} (D);
			\end{tikzpicture}
		\end{adjustbox}
		\caption{}
		\label{fig:right-io}
	\end{subfigure}
	\label{fig:io}
	\caption{(a) $\aut$, $\lang(\aut) = \{ddd, abc, bc\}$. (b) $\aut'$, $\lang(\aut') = \{bcd, aaab\}$}
\end{figure}
%
%The additive lift of the \code{indexOf} operation is $\indexofs{\lin}{\lin'} = \sset{\indexofs{\sigma}{\sigma'}}{\sigma \in \lin, \sigma\in \lin'}$. 
The abstract semantics of \code{indexOf} is $\mathsf{IO}^\sharp:\fa \times \fa \rightarrow \interval$
and should return the interval of any possible positions of strings in a language inside strings of another language.
Consider for instance the automaton $\aut$ in Fig.~\ref{fig:left-io} and suppose to call 
$\indexofa{\aut}{\aut'}$ where $\aut' = \minimize(\{bc\})$. 
The idea is that of building, for each state $q$ in $\aut$, the automaton $\aut_{q}$ which is $\aut$ where all the states are final and the initial state is $q$. Hence, we check whether 
%$\aut_{q}$ accepts strings accepted by $\aut'$ ($bc$), i.e., we check non-emptiness of 
$\aut_{q} \sqcap \aut'$ is non empty and we collect the size of the maximum path from $q_0$ to $q$ in $\aut$.
% In the example, $bc$ can be found both at index 1 and 2. 
If there exists at least one state from which any string accepted by $\aut'$ cannot be read, we collect -1. In the example, $\aut_{q_0}$ adds $\{0\}$,  $\aut_{q_1}$ adds $\{1\}$, while all the other states add $\{-1\}$. 
Finally, we return the interval $[\mathsf{min}\{-1,1,0\}, \mathsf{max}\{-1, 1,0\}] = [-1, 1]$. The full algorithm is reported in Alg.~\ref{alg:indexof}. 
%Lines 2-12 search for the indexes of $\aut$ from which can be found a string of $\aut'$, as explain before. If no index is found, -1 is added to $\mathsf{indexesOf}$ (line 10). Otherwise, we add the length of the minimum path from $q_0$ to $q$ and then we check if there exists a path with cycle from $q_0$ to $q$: in this case, we cannot finitely determinate the result (i.e., return a finite interval) since a path to reach $q$ contains a cycle. Hence, we add $+\infty$ to $\mathsf{indexesOf}$\footnote{If $X$ contains $+\infty$, then $\mathsf{max}(X)=+\infty$}. In the case of both $\aut$ and $\aut'$ recognize exactly one string we can precisely answer, returning the interval corresponding to the minimum index found (line 13-14). Otherwise, the result is the interval from the minimum index to the maximum index. Note that, if from any  state of $\aut$ a string of $\aut'$ can be read, the interval $[-1,-1]$ is returned.
\begin{theorem} \label{thm:io}
	$\mathsf{IO}^\sharp$ is sound but not complete: $\forall \aut,\aut' \in \fa\:.$
	%
	%\[
	$\indexofs{\lang(\aut)}{\lang(\aut')} \subset \indexofa{\aut}{\aut'}
	$. 
	%where $\indexofs{\lang(\aut)}{\lang(\aut')}$ is the lift of the concrete semantics to languages.
\end{theorem}

As a counterexample to completeness, consider the automaton $\aut'$ in Fig.\ref{fig:right-io} and the automaton $\aut'' = \minimize(\{b\})$: $\indexofa{\aut'}{\aut''}=[-1, 3]$ $\not\subset \indexofs{\lang(\aut')}{\lang(\aut'')} = \{0, 3\}$. The interval $[-1, 3]$ contains also indexes where the string $b$ is not recognized (e.g., 2), but it also contains the information ($-1$) meaning that there exists at least one accepted string without $b$ as substring, which is not true.

%\begin{figure}[t]
%	\scalebox{0.85}{%
%	\begin{algorithm}[H]
%		\KwData{$\aut = (Q, \Sigma, \delta, q_0, F), \aut'=(Q', \Sigma, \delta', q_0', F')$}
%		\KwResult{$\indexofa{\aut}{\aut'}$}
%		
%		$\mathsf{indexesOf} \leftarrow \varnothing$
%	
%		\ForEach{$q \in Q$}{
%			$\aut_{q} \leftarrow (Q, \Sigma, \delta, q, Q)$\;
%			\uIf{$\aut_{q} \glbfa \aut' \neq \varnothing$}{
%				$\mathsf{indexesOf} \leftarrow \mathsf{indexesOf} \cup \{\mathsf{len}(\mathsf{minPath}(\aut, q_0, q))\}$\;
%				
%				\lIf{$\exists p=\mathsf{path}(q_0,q)\mbox{ s.t. } \mathsf{hasCycle}(p)$}{
%					$\mathsf{indexesOf} \leftarrow \mathsf{indexesOf} \cup \{+\infty\}$
%				}
%			}	
%		
%			\Else{
%				$\mathsf{indexesOf} \leftarrow \mathsf{indexesOf} \cup \{-1\}$\;
%			}	
%		}
%		
%		\uIf{$|\lang(\aut)| == |\lang(\aut')| == 1$}{
%				\textbf{return} $[\mathsf{min}(\mathsf{indexesOf}), \mathsf{min}(\mathsf{indexesOf})]$\;
%			}
%		\Else{
%				\textbf{return} $[\mathsf{min}(\mathsf{indexesOf}), \mathsf{max}(\mathsf{indexesOf})]$\;
%			}
%		\caption{$\mathsf{IO}^\sharp: \fa \times \fa \rightarrow \interval$ algorithm}
%		\label{alg:indexof}
%	\end{algorithm}
%}
%\end{figure}
\begin{figure}[t]
\begin{minipage}[t]{0.48\textwidth}
\vspace{0pt}  	
\scalebox{0.82}{%
		\begin{algorithm}[H]
		\KwData{$\aut = (Q, \Sigma, \delta, q_0, F)$} 
		\KwResult{$\lengtha{\aut}$}
		$\mathsf{P\_len} \leftarrow 0$;	$\mathsf{p\_len} \leftarrow \infty $ 
		
		\eIf{$\mathsf{hasCycle}(\aut)$}{
			\ForEach{$q_f \in F$} {
				$\mathsf{p} \leftarrow \mathsf{minPath}(\aut, q_0, q_f)$\;		
				
				\lIf{$\mathsf{len}(\mathsf{p}) < \mathsf{p\_len}$}{
					$\mathsf{p\_len} \leftarrow \mathsf{len}(\mathsf{p})$
				}	
			}		
			\textbf{return} $[\mathsf{p\_len}, +\infty]$\;			
		}
		{
			\ForEach{$q_f \in F$} {
				$\mathsf{p} \leftarrow \mathsf{minPath}(\aut, q_0, q_f)$\;
				$\mathsf{P} \leftarrow \mathsf{maxPath}(\aut, q_0, q_f)$\;
				
				\lIf{$\mathsf{len}(\mathsf{p}) < \mathsf{p\_len}$}{
					$\mathsf{p\_len} \leftarrow \mathsf{len}(\mathsf{p})$
				}	
				
				\lIf{$\mathsf{len}(\mathsf{P}) > \mathsf{P\_len}$}{
					$\mathsf{P\_len} \leftarrow \mathsf{len}(\mathsf{P})$
				}
			}
			\textbf{return} $[\mathsf{p\_len}, \mathsf{P\_len}]$\;			
		}
		
		\caption{$\mathsf{LE}^\sharp: \fa \rightarrow \interval$ alg.}
		\label{alg:length}
	\end{algorithm}
}
\end{minipage}%
\begin{minipage}[t]{0.60\textwidth}
\vspace{0pt}
	\scalebox{0.82}{%
			\begin{algorithm}[H]
			\KwData{$\aut = (Q, \Sigma, \delta, q_0, F), \aut'=(Q', \Sigma, \delta', q_0', F')$}
			\KwResult{$\indexofa{\aut}{\aut'}$}
			
			$\mathsf{indexesOf} \leftarrow \varnothing$
			
			\ForEach{$q \in Q$}{
				$\aut_{q} \leftarrow (Q, \Sigma, \delta, q, Q)$\;
				\uIf{$\aut_{q} \glbfa \aut' \neq \varnothing$}{
					$\mathsf{indexesOf} \leftarrow \mathsf{indexesOf} \cup \{\mathsf{len}(\mathsf{maxPath}(\aut, q_0, q))\}$\;
					
					\If{$\exists p=\mathsf{path}(q_0,q)\mbox{ s.t. } \mathsf{hasCycle}(p)$}{
						$\mathsf{indexesOf} \leftarrow \mathsf{indexesOf} \cup \{+\infty\}$
					}
				}	
				
				\Else{
					$\mathsf{indexesOf} \leftarrow \mathsf{indexesOf} \cup \{-1\}$\;
				}	
			}
			
			\uIf{$|\lang(\aut)| == |\lang(\aut')| == 1$}{
				\textbf{return} $[\mathsf{min}(\mathsf{indexesOf}), \mathsf{min}(\mathsf{indexesOf})]$\;
			}
			\Else{
				\textbf{return} $[\mathsf{min}(\mathsf{indexesOf}), \mathsf{max}(\mathsf{indexesOf})]$\;
			}
			\caption{$\mathsf{IO}^\sharp: \fa \times \fa \rightarrow \interval$ alg.}
			\label{alg:indexof}
		\end{algorithm}
}
\end{minipage}
\end{figure}
\subsection{Abstract semantics of concatenation}
The abstract semantics of string concatenation is $\mathsf{CC}^\sharp:\fa \times \fa \rightarrow \fa$ and returns the concatenation between two automata. Since regular languages are closed under concatenation, the property also holds on automata. In Fig.~\ref{fig:conc}, we report an example of concatenation between two automata. Hence, $\mathsf{CC}^\sharp$ exactly implements the standard concatenation operation between automata. Given the closure property on automata, the following result holds. 

\begin{theorem} \label{thm:concat} $\mathsf{CC}^\sharp$ is sound and complete: $\forall \aut,\aut' \in \fa\:.\concs{\lang(\aut)}{\lang(\aut')} = \conca{\aut}{\aut'}.$
\end{theorem}

\begin{figure}[t]
	\centering
	(a)\begin{subfigure}[t]{0.33\textwidth}
		\centering
		\begin{adjustbox}{scale=0.75} 
			\begin{tikzpicture}[->,>=stealth',shorten >=1pt,auto,node distance=1.9cm, semithick]
				\node[initial,state,scale=\nodesize, initial text =] 				 (A)                    {};
				\node[state,scale=\nodesize, accepting]      			 			 (B) [above right of=A] 		{};
				\node[state,scale=\nodesize, accepting]        			 	  (C) [ right of=A] 		{};

				\path[->] (A) edge node {$a$} (B);		
				\path[->] (B) edge [loop above] node {$a$} (B);
				\path[->] (A) edge node[swap] {$b$} (C);
			\end{tikzpicture}
		\end{adjustbox}
		\caption{}
		\label{fig:left-conc}
	\end{subfigure}%
	~ 
	(b)\begin{subfigure}[t]{0.3\textwidth}
		\centering
		\begin{adjustbox}{scale=0.75} 
			\begin{tikzpicture}[->,>=stealth',shorten >=1pt,auto,node distance=1.9cm, semithick]
			\node[initial,state,scale=\nodesize, initial text =] 				 (A)                    {};
			\node[state,scale=\nodesize, accepting]      			 					 (B) [right of=A] 		{};
			
			\path[->] (A) edge node {$c$} (B);
			\path[->] (B) edge [loop above] node {$d$} (B);
			\end{tikzpicture}
		\end{adjustbox}
		\caption{}
		\label{fig:right-conc}
	\end{subfigure}
	~
	(c)\begin{subfigure}[t]{0.2\textwidth}
		\centering
		\begin{adjustbox}{scale=0.75} 
			\begin{tikzpicture}[->,>=stealth',shorten >=1pt,auto,node distance=1.9cm, semithick]
			\node[initial,state,scale=\nodesize, initial text =] 				 (A)                    {};
			\node[state,scale=\nodesize]      			 					 (B) [above right of=A] 		{};
			\node[state,scale=\nodesize]        			 				     (C) [right of=A] 		{};		
			\node[state,scale=\nodesize, accepting]        			 				     (D) [ right of=C] 		{};
			%\node[state,scale=\nodesize, accepting]        			 				     (E) [right of=D]   {};
	
			\path[->] (A) edge node {$a$} (B);		
			\path[->] (B) edge [loop above] node {$a$} (B);
			\path[->] (A) edge node[swap] {$b$} (C);		
			\path[->] (B) edge node {$c$} (D);
			\path[->] (C) edge node[swap] {$c$} (D);
%			\path[->] (D) edge node {$d$} (E);
			\path[->] (D) edge [loop above] node {$d$} (D);
			\end{tikzpicture}
		\end{adjustbox}
		\caption{}
		\label{fig:conc-bot}
	\end{subfigure}
	\caption{(a) $\aut$, $\lang(\aut) = \sset{a^n}{n > 0} \cup \{b\}$ (b) $\aut'$, $\lang(\aut') = \sset{cd^n}{n \in \mathbb{N}}$ (c) $\aut'' = \conca{\aut}{\aut'}$}
	\label{fig:conc}
\end{figure}
\subsection{Concerning abstract implicit type conversion}
In this section, we discuss the abstraction of the implicit type conversion functions. For space limitations, we will focus only on the conversion of automata into other values, since the conversions concerning booleans, not-a-number and intervals are standard. Let $\code{toBool}^\sharp: \aval \rightarrow \abool$ be applied to $\aut \in \fa$: If $\aut \sqcap \minimize(\{\epsilon\})  = \varnothing$, it returns $\{\true\}$, when $\aut = \minimize(\{\epsilon\})$ the function returns $\{\false\}$, otherwise %$\aut$ contains  strings that can be converted both to $\true$ and to $\false$ and 
the function returns $\{\true, \false\}$. Implicit type conversion to $\fa$ is handled by the function $\code{toStr}^\sharp: \aval \rightarrow \fa$. As far as non numeric strings are concerned, $\code{toStr}^\sharp$ returns $\minimize(\{\code{NaN}\})$. If the input is the boolean value $\true$ [$\false$] it returns
$\minimize(\{\true\})$ [$\minimize(\{\false\})$], otherwise it returns $\minimize(\{\true\}) \sqcup \minimize(\{\false\})$. Converting intervals to FA is more tricky. If $l,h \in \mathbb{Z}$, the conversion to automata is simply $\bigsqcup_{i \in [l,h]} \minimize(\{\mathcal{S}(i)\})$. 
\begin{figure}[t]
	(a)\begin{subfigure}[t]{0.5\textwidth}
		\centering	
		\begin{adjustbox}{scale=0.75} 
			\begin{tikzpicture}[->,>=stealth',shorten >=1pt,auto,node distance=3.5cm, semithick]
			\node[initial,state,scale=0.45, initial text =] 	 (A)                    {};
			\node[state,scale=0.45]      			 			 (B) [right of=A] 		{};
			\node[state,scale=0.45,accepting]        			 		 (C) [right of=B] 		{};
			
			\path[->] (A) edge node {+} (B);
			\path[->] (B) edge node {$[0-9]$} (C);
			\path[->] (C) edge [loop above] node {$[0-9]$} (C);
			\path[->] (A) edge [bend right] node[swap] {$[0-9]$} (C);
			\end{tikzpicture}
		\end{adjustbox}
		\caption{}
		\label{fig:ty-plus}
	\end{subfigure}%
	~ 
	(b)\begin{subfigure}[t]{0.5\textwidth}
		\centering	
		\begin{adjustbox}{scale=0.75} 
			\begin{tikzpicture}[->,>=stealth',shorten >=1pt,auto,node distance=3.5cm, semithick]
			\node[initial,state,scale=\nodesize, initial text =] 	 (A)                    {};
			\node[state,scale=\nodesize]      			 			 (B) [right of=A] 		{};
			\node[state,scale=\nodesize,accepting]        			 		 (C) [right of=B] 		{};
	
			\path[->] (A) edge node {-} (B);
			\path[->] (B) edge node {$[0-9]$} (C);
			\path[->] (C) edge [loop above] node {$[0-9]$} (C);
			\path[->] (A) edge [bend right, draw=none] node[swap] {} (C);
			\end{tikzpicture}
			\end{adjustbox}
		\caption{}
		\label{fig:ty-minus}
	\end{subfigure}
	\caption{(a) $\mathsf{toStr}^\sharp([0, +\infty])$. (b) $\mathsf{toStr}^\sharp([-\infty, 0])$}
	\label{fig:atostr}
\end{figure}
The interval-to-automaton conversion for $[0, +\infty]$ and $[-\infty, 0]$ are respectively shown in Fig.~\ref{fig:ty-plus} and Fig.~\ref{fig:ty-minus}. Other unbounded intervals, $[+l, +\infty]$ and $[-l, +\infty]$ ($l > 0$), are converted in $\atostr{[0, +\infty]} \smallsetminus \atostr{[0, l]}$ and $\atostr{[-l, 0]} \sqcup \atostr{[0, +\infty]}$, respectively. Conversions of intervals $[-\infty, l]$ and $[-\infty, -l]$ ($l > 0$) are analogous, %while the conversion of $[+\infty, -\infty]$ is $\atostr{[-\infty, 0]} \sqcup \atostr{[0, +\infty]}$. 
while, $\atostr{[-\infty, +\infty]} = \minimize(\Sigma_{\mathbb{Z}})$. Finally, $\mathsf{toInt}^\sharp: \aval \rightarrow \interval \cup \{\anan\}$ handles conversion to intervals. 
Given an automaton $\aut$, 
%the conversion to interval has to check if there exists a numeric string recognized by $\aut$. I
if $\aut \sqcap \minimize(\Sigma_{\mathbb{Z}}) = \varnothing$, the automaton is precisely converted to $\cnan$, otherwise, if $\aut \leqfa \minimize(\Sigma_{\mathbb{Z}})$ it means that $\lang(\aut)$ contains only numeric strings. In this case, if $\aut$ accepts a finite number of strings, we convert each $\sigma \in \lang(\aut)$ to the corresponding number and return the interval from the minimum to the maximum number. In the other cases, we check whether $\aut$ recognizes positive numeric strings (checking if the initial state reads only $+$ or number symbols), negative numeric strings (checking if the initial state reads only $-$ or 0 symbols) or both. In the first case, we return $[0, +\infty]$, in the second $[-\infty, 0]$ and in the last $[-\infty, +\infty]$.

The abstract interpreter for the abstract semantics so far defined has been tested by means of the implementation of an automata library\footnote{Available at \url{www.github.com/SPY-Lab/fsa} and the $\mujs$ static analyzer at \url{www.github.com/SPY-Lab/mu-js}}. This library includes the implementation of all the algorithms concerning the finite state automata domain and provide well-known operations on automata such as suffix, right quotient, and abstract domain-related operations, such as $\lubfa$, $\glbfa$, and a parametric widening for tuning precision and forcing convergence. The library is suitable and easily pluggable into existing static analyzers, such as \cite{safe, tajs2009, safe2012, jsai2014}. The bottleneck of our library is the determinization operation, having exponential complexity \cite{hopcroft1979} (we rely on determinization in the minimization algorithm, in order to preserve the automata arising during the abstract computations minimum and deterministic). 
It is worth noting that, as reported in Thm.~\ref{thm:fa-moore-family}, $\wp(\Sigma^*)$ (string concrete domain) and $\fa$ (abstract string domain) do not form a Galois connection but, nevertheless, this is not a concern. We have shown, for the core language we adopted, that the abstract semantics we have defined for string operations guarantee soundness hence, if the abstract interpreter starts from regular initial conditions (i.e., constraints expressible as finite state automata) it will always compute regular invariants. Indeed, it is sound to start from $\top$ initial condition that, in our string abstract domain, is expressible by $\minimize(\wp(\Sigma^*))$, which is regular.

\begin{figure}[t]
	\centering
	\includegraphics[scale=0.75]{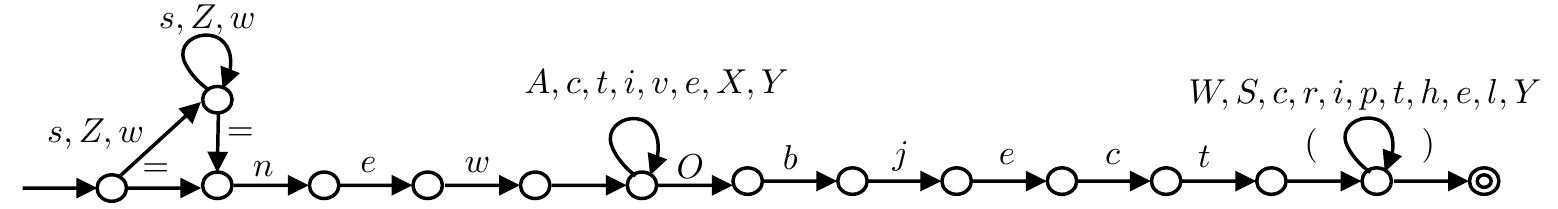}
	\caption{$\aut_{\scriptsize{d}}$ abstract value of \code{d} before \code{eval} call of the program in Fig.~\ref{fig:mal}}
	\label{fig:faut}
\end{figure}

%We have implemented an automaton library\footnote{Available at \url{https://github.com/SPY-Lab/finite-state-automata}} providing well-known operations on automata such as suffix, right/left quotient, and abstract domain-related operations, such as $\lubfa, \glbfa, \nabla_n$, and the algorithms proposed in the paper, making the library suitable and easily pluggable into a generic abstract interpreter for string analysis. The bottleneck of our library is the determinization operation, having exponential complexity \cite{hopcroft1979} (we rely on determinization in the minimization algorithm, in order to keep the automata arising during the abstract computations minimum and deterministic).
%
\noindent
\textbf{Example: Obfuscated malware.} Consider the fragment reported in Fig.~\ref{fig:mal} in the introduction. 
By computing the abstract semantics of this code, we obtain that the abstract value of \code{d}, at the \code{eval} call, is the automaton $\aut_{\scriptsize{d}}$ in Fig.~\ref{fig:faut}. The cycles are caused by the widening application in the \code{while} computation. 
From this automaton we are able to retrieve some important and non-trivial information. For example, we are able to answer to the following question: \emph{May $\aut_{\scriptsize{d}}$ contain a string corresponding to an assignment to an ActiveXObject?} We can simply answer by checking the predicate $\aut_{\scriptsize{d}} \sqcap \minimize(\id \cdot \{new \ ActiveXObject(\} \cdot \Sigma^* \cdot \{)\}) \neq \varnothing$, checking whether $\aut_{\scriptsize{d}}$ recognizes strings that are concatenations of any identifier with the string $new \ ActiveXObject$, followed by  any possible string. In the example, the predicate returns $\true$. Another interesting  information could be: \emph{May $\aut_{\scriptsize{d}}$ contain \code{eval} string?} We can answer by checking whether $\aut_{\scriptsize{d}} \sqcap \minimize(\{eval\}) \neq \varnothing$, that is false and guarantees that no explicit call to \code{eval} can occur. \\
We observe that such analysis may lose precision during fix-point computations, causing the cycles in the automaton in Fig.~\ref{fig:faut}, due to the widening application. Nevertheless, it is worth noting that this result is obtained without any precision improvement on fix-point computations, such as loop unrolling or widening with thresholds. We think these analyses will drastically decrease false positives of the proposed string analysis but we will address this topic in future~work. 

\section{Discussion and related work}\label{sect:rel}

In this paper, we have proposed an abstract semantics for a toy imperative language $\mujs$, augmented with string manipulation operations, expressive enough to handle dynamic typing and implicit type conversion. In our abstract semantics, we have combined the DFA domain with abstract domains for the other primitive types, necessary to deal with static analysis of programs with dynamic typing. The proposed framework allows us to formally prove soundness and to study precision of the abstract semantics of each string operation: Depending on the property of interest, one can tune the degree of precision, namely the completeness of any string~operation. 

%\noindent
%\textbf{Analysis vs verification.}
%Even if several solutions, also involving finite state machines, have been proposed for string solving and verification\cite{yu2008, norn, smt-string}, it is worth noting that our approach is placed instead in the context of string static analysis. Among the years, there has always been the intuition that program analysis was \textit{harder} than verification: given a program, the aim of the former is to derive invariants for each program point, the one of the latter is instead to check whether a certain property holds for the given input program. Recently, this concept has been formalized from a computability point of view\cite{cousot2018}, confirming this belief. Hence, our approach, placed in the context of static analysis of string manipulation programs, has goals that are hardly comparable with the solutions proposed in  the context of verification, such those~cited~above.

\noindent
\textbf{Main related work.} The issue of analyzing strings is a widely studied problem, and it has been tackled in the literature from different points of view. Before discussing the most related works, we can observe what makes our approach original w.r.t.  all the existing ones: (1) We provide a modular abstract domain parametric on the the abstractions of the different primitive types, this allows us both to obtain a tunable semantics precision and 
to handle dynamic typing for operation having both integer and string parameters, e.g., \code{substring}; (2) Our focus is on the characterization of a formal abstract interpretation-based framework where it is possible to prove  soundness and to analyze completeness of string operations, in order to understand where it is possible to tune precision versus efficiency. \\
The main feature we have in common with existing works is the use of DFA (regular expressions) for abstracting strings.
%Nevertheless, several existing works have involved finite state automata  and regular expressions in different contexts (and different ways) of static analysis. In this section, we report some of the paper that most has followed our approach and inspired our work.
In \cite{yu2008}, the authors propose symbolic string verifier for PHP based on finite state automata, represented by a particular form of binary decision diagrams, the MBDD. Even if it could be interesting to understand whether this representation of DFAs may be used also for improving our algorithms, their work only considers operations exclusively involving strings (not also integers such as \code{substring}) and therefore it provides a solution for different string manipulations.
%
%the authors propose symbolic string analysis for PHP based on finite state automata, represented by a particular form of binary decision diagrams, the MBDD. The authors focus on defining three automata operations: concatenation, replacement and closure. Since their representation choice, that does not provide $\epsilon$-transitions, and the potentially non-determinism introduced by the considered operations, the authors needed to extends the alphabet of DFAs to handle $\epsilon$-transitions, to perform the cited string operations and finally to project the resulting DFA to the original alphabet.
%From the other side, our work differs from \cite{yu2008} since we formalize the finite state automata  domain in the context of abstract interpretation, showing the non-existence of a Galois connection between $\wp(\Sigma^*)$ and $\fa$ and proving the closure of other four string operations that also involve other type values. Hence, we do not need any alphabet extension to handle potential non-determinism introduced by the abstract semantics of the proposed string operations: starting from regular initial constraints, the string-value analysis will be always return a regular constraint.
%Moreover, we have recast the finite state automata domain in a more complex abstract domain including other type abstract domain. This choice was guided by the fact that the string operations we have studied rely also on other type values and not only on strings (as in \cite{yu2008}) and combination of different abstract domains turned out to be necessary. 
In \cite{wid-approach}, the authors propose an abstract interpretation-based string analyzer approximating strings into a subset of regular languages, called \textit{regular strings} and they define the abstract semantics for four string operations of interest together with a widening. This is the most related work, but our approach is strictly more general, since we do not introduce any restriction of regular languages and we abstract integers on intervals instead of on constants (meaning that our domain is strictly more precise). 
%
%\code{substring}, \code{trim}, \code{replace} and concatenation and a widening for the regular strings domain to accelerate fix-point computations. Since the abstract domain restrictions of \cite{wid-approach}, our abstract domain of string is more expressive, since we can precisely express  any regular language.
In \cite{park2016}, the authors propose a  scalable static analysis for jQuery that relies on a novel abstract domain of regular expressions. The abstract domain in \cite{park2016} contains the finite state automata domain but pursues a different task and does not provide semantics for string manipulations. Surely it may be interesting to integrate our library for string manipulation operators into SAFE. 
%
%Nevertheless, the aim of our work is to define a precise abstract semantics of string manipulations programs. In future work, it may be interesting to plug in our abstract semantics into SAFE static analyzer and discuss and studying performances and precisions between the two choices. 
Finally, \cite{midtgaard16} proposes a generalization of regular expression, formally illustrating a parametric abstract domain of regular expressions starting from a complete lattice of reference. However, this work does not tackle the problem of analyzing string manipulations, since it instantiates the parametric abstract domain in the network communication environment, analyzing the exchanged messages~as regular~expressions.\\
Finite state machines (transducer and automata) have found a critical application also in model checking both for enforcing string constraints and to model infinite transition systems \cite{popl2016}. For example, the authors of \cite{cav2014} define a sound decision procedure for a  regular language-based logic for  verification of string properties. The authors of \cite{abs-reg-mc} propose an automata abstraction in the context of regular model checking to tackle the well-known problem of state space explosion. 
Moreover, other formal systems, similar to DFA, have been proposed in the context of string analysis \cite{tree-automata,visibly, afa}. As future work, it can be interesting to study the relation between standard DFA and the other existing formal models, such as logics or other forms of FA. An interesting recent work is reported in~\cite{olliaro2018}, where the authors propose M-String, a parametric string abstract domain that extends the segmentation approach proposed in~\cite{logozzo2011} for C strings. M-String uses an abstract domain for the content of a string and an abstract domain for expression, inferring when a string index position corresponds to an expression of the considered abstract domain. As future work, it could be interesting studying how to involve the finite state automata abstract domain into M-String, as abstraction of the string content. \\
%%% JavaScript string static analysis related work
In the context of JavaScript, several static analyzers have been proposed, pushed by the wide range of applications and the security issues related to the language\cite{safe, tajs2009, safe2012, jsai2014}. TAJS\cite{tajs2009} is a static analyzer based on abstract interpretation for JavaScript. The authors focus on allocation site abstraction, plugging in the static analyzer the \textit{recency abstraction}\cite{recency}, decreasing the number of false positives when objects are accessed. Upon TAJS, the authors have defined a sound way to statically analyze a large range of non-trivial \code{eval} patterns \cite{evil}. In \cite{safe}, the authors define the Loop-Sensitive Analysis (LSA) that distinguishes loop iterations using \textit{loop strings}, in the same way \textit{call strings} distinguish function calls from different call sites in $k$-CFA\cite{cfa}. The authors have implemented LSA into SAFE\cite{safe2012}, a JavaScript web applications static analyzer. As future work, it may be interesting to combine LSA with our abstract semantics for decreasing the false positives introduced by the widening during fix-point computations. Finally, in \cite{amadini17}, the authors extend SAFE defining a formal framework to combine multiple existing string abstract domains for the analysis of JavaScript programs, showing that combinations of simple abstract domains out-perform the precision of the existing state-of-art static analyzers, comparing their approach with SAFE, TAJS and JSAI. \\
\noindent
\textbf{Future ideas.}
In this paper we have proposed string static program analysis for a set of relevant string manipulation operations, whose semantics is inspired by the JavaScript behaviors. We are currently working on extending our framework in order to fully cover the JavaScript \code{String} built-in global  object, formally defining the remaining methods contained in it. Afterwards, the first goal is to integrate our abstract semantics into a static analyzer for JavaScript, that uses finite state automata to approximate strings. In order to decrease the number of false positives in our string approximation in presence of loops, several techniques will be involved, such as loop unrolling and LSA\cite{safe}. The domain described in this paper has been equipped only with a widening, to enforce termination in fix-point computations, that may lead to a big loss of precision. A narrowing will be studied and introduced in our static analyzer in order to retrieve some precision lost when widening is~applied. \\
We conclude by observing that we are strongly confident that an important future application of our semantics may be the string-to-code primitives analysis. Consider, for instance, in JavaScript programs, the \code{eval} function, transforming strings into code. As already observed, our semantics is sound and precise enough for answering to some non-trivial property of interest. Hence, we think this semantics for strings can be a good starting point for a sound and {\em precise enough} analysis of \code{eval}, for example in JavaScript, which is still an open problem in static analysis. 
%0uy43 
%Finally, as mentioned before, we think that such (collecting) abstract domain for strings plus the defined abstract semantics will be helpful to define  a sound and precise analysis for handling string-to-code primitives, such as \code{eval} in JavaScript or Python. 

%\section{Conclusion}\label{sect:concl}
%\input{conclusion}

\bibliographystyle{eptcs}
\bibliography{biblio2}

\begin{thebibliography}{10}
\providecommand{\bibitemdeclare}[2]{}
\providecommand{\surnamestart}{}
\providecommand{\surnameend}{}
\providecommand{\urlprefix}{Available at }
\providecommand{\url}[1]{\texttt{#1}}
\providecommand{\href}[2]{\texttt{#2}}
\providecommand{\urlalt}[2]{\href{#1}{#2}}
\providecommand{\doi}[1]{doi:\urlalt{http://dx.doi.org/#1}{#1}}
\providecommand{\bibinfo}[2]{#2}

\bibitemdeclare{inproceedings}{cav2014}
\bibitem{cav2014}
\bibinfo{author}{P.~\surnamestart Abdulla\surnameend},
  \bibinfo{author}{M.~\surnamestart Atig\surnameend},
  \bibinfo{author}{Y.~\surnamestart Chen\surnameend},
  \bibinfo{author}{L.~\surnamestart Hol{\'{\i}}k\surnameend},
  \bibinfo{author}{A.~\surnamestart Rezine\surnameend},
  \bibinfo{author}{P.~\surnamestart R{\"{u}}mmer\surnameend} \&
  \bibinfo{author}{J.~\surnamestart Stenman\surnameend} (\bibinfo{year}{2014}):
  \emph{\bibinfo{title}{String {C}onstraints for {V}erification}}.
\newblock In: {\sl \bibinfo{booktitle}{{CAV}'14}},
  \doi{10.1007/978-3-319-08867-9\_10}.

\bibitemdeclare{inproceedings}{visibly}
\bibitem{visibly}
\bibinfo{author}{R.~\surnamestart Alur\surnameend} \&
  \bibinfo{author}{P.~\surnamestart Madhusudan\surnameend}
  (\bibinfo{year}{2004}): \emph{\bibinfo{title}{Visibly pushdown languages}}.
\newblock In: {\sl \bibinfo{booktitle}{{STOC}'04}},
  \doi{10.1145/1007352.1007390}.

\bibitemdeclare{inproceedings}{amadini17}
\bibitem{amadini17}
\bibinfo{author}{R.~\surnamestart Amadini\surnameend},
  \bibinfo{author}{A.~\surnamestart Jordan\surnameend},
  \bibinfo{author}{G.~\surnamestart Gange\surnameend},
  \bibinfo{author}{F.~\surnamestart Gauthier\surnameend},
  \bibinfo{author}{P.~\surnamestart Schachte\surnameend},
  \bibinfo{author}{H.~\surnamestart S{\o}ndergaard\surnameend},
  \bibinfo{author}{P.~J. \surnamestart Stuckey\surnameend} \&
  \bibinfo{author}{C.~\surnamestart Zhang\surnameend} (\bibinfo{year}{2017}):
  \emph{\bibinfo{title}{Combining String Abstract Domains for JavaScript
  Analysis: An Evaluation}}.
\newblock In: {\sl \bibinfo{booktitle}{{TACAS}'17}},
  \doi{10.1007/978-3-662-54577-5\_3}.

\bibitemdeclare{article}{arceri2017}
\bibitem{arceri2017}
\bibinfo{author}{V.~\surnamestart Arceri\surnameend} \&
  \bibinfo{author}{S.~\surnamestart Maffeis\surnameend} (\bibinfo{year}{2017}):
  \emph{\bibinfo{title}{{A}bstract {D}omains for {T}ype {J}uggling}}.
\newblock {\sl \bibinfo{journal}{ENTCS}} \bibinfo{volume}{331},
  \doi{10.1016/j.entcs.2017.02.003}.

\bibitemdeclare{inproceedings}{recency}
\bibitem{recency}
\bibinfo{author}{G.~\surnamestart Balakrishnan\surnameend} \&
  \bibinfo{author}{T.~\surnamestart Reps\surnameend} (\bibinfo{year}{2006}):
  \emph{\bibinfo{title}{Recency-{A}bstraction for {H}eap-{A}llocated
  {S}torage}}.
\newblock In: {\sl \bibinfo{booktitle}{{SAS}'06}}, \doi{10.1007/11823230\_15}.

\bibitemdeclare{inproceedings}{DBLP:conf/cav/BartzisB04}
\bibitem{DBLP:conf/cav/BartzisB04}
\bibinfo{author}{C.~\surnamestart Bartzis\surnameend} \&
  \bibinfo{author}{T.~\surnamestart Bultan\surnameend} (\bibinfo{year}{2004}):
  \emph{\bibinfo{title}{{W}idening {A}rithmetic {A}utomata}}.
\newblock In: {\sl \bibinfo{booktitle}{{CAV}'04}},
  \doi{10.1007/978-3-540-27813-9\_25}.

\bibitemdeclare{article}{bordihn09}
\bibitem{bordihn09}
\bibinfo{author}{H.~\surnamestart Bordihn\surnameend},
  \bibinfo{author}{M.~\surnamestart Holzer\surnameend} \&
  \bibinfo{author}{M.~\surnamestart Kutrib\surnameend} (\bibinfo{year}{2009}):
  \emph{\bibinfo{title}{Determination of finite automata accepting subregular
  languages}}.
\newblock {\sl \bibinfo{journal}{Theor. Comput. Sci.}}
  \bibinfo{volume}{410}(\bibinfo{number}{35}), \doi{10.1016/j.tcs.2009.05.019}.

\bibitemdeclare{inproceedings}{tree-automata}
\bibitem{tree-automata}
\bibinfo{author}{A.~\surnamestart Bouajjani\surnameend},
  \bibinfo{author}{P.~\surnamestart Habermehl\surnameend},
  \bibinfo{author}{L.~\surnamestart Hol{\'{\i}}k\surnameend},
  \bibinfo{author}{T.~\surnamestart Touili\surnameend} \&
  \bibinfo{author}{T.~\surnamestart Vojnar\surnameend} (\bibinfo{year}{2008}):
  \emph{\bibinfo{title}{Antichain-{B}ased {U}niversality and {I}nclusion
  {T}esting over {N}ondeterministic {F}inite {T}ree {A}utomata}}.
\newblock In: {\sl \bibinfo{booktitle}{{CIAA}'08}},
  \doi{10.1007/978-3-540-70844-5\_7}.

\bibitemdeclare{inproceedings}{abs-reg-mc}
\bibitem{abs-reg-mc}
\bibinfo{author}{A.~\surnamestart Bouajjani\surnameend},
  \bibinfo{author}{P.~\surnamestart Habermehl\surnameend} \&
  \bibinfo{author}{T.~\surnamestart Vojnar\surnameend} (\bibinfo{year}{2004}):
  \emph{\bibinfo{title}{Abstract {R}egular {M}odel {C}hecking}}.
\newblock In: {\sl \bibinfo{booktitle}{{CAV}'04}},
  \doi{10.1007/978-3-540-27813-9\_29}.

\bibitemdeclare{article}{campeanu2002}
\bibitem{campeanu2002}
\bibinfo{author}{C.~\surnamestart C{\^{a}}mpeanu\surnameend},
  \bibinfo{author}{A.~\surnamestart Paun\surnameend} \&
  \bibinfo{author}{S.~\surnamestart Yu\surnameend} (\bibinfo{year}{2002}):
  \emph{\bibinfo{title}{An {E}fficient {A}lgorithm for {C}onstructing {M}inimal
  {C}over {A}utomata for {F}inite {L}anguages}}.
\newblock {\sl \bibinfo{journal}{Int. J. Found. Comput. Sci.}}
  \bibinfo{volume}{13}(\bibinfo{number}{1}), \doi{10.1142/S0129054102000960}.

\bibitemdeclare{inproceedings}{wid-approach}
\bibitem{wid-approach}
\bibinfo{author}{T.~\surnamestart Choi\surnameend},
  \bibinfo{author}{O.~\surnamestart Lee\surnameend},
  \bibinfo{author}{H.~\surnamestart Kim\surnameend} \&
  \bibinfo{author}{K.~\surnamestart Doh\surnameend} (\bibinfo{year}{2006}):
  \emph{\bibinfo{title}{A {P}ractical {S}tring {A}nalyzer by the {W}idening
  {A}pproach}}.
\newblock In: {\sl \bibinfo{booktitle}{{APLAS}'06}},
  \doi{10.1007/11924661\_23}.

\bibitemdeclare{inproceedings}{olliaro2018}
\bibitem{olliaro2018}
\bibinfo{author}{Agostino \surnamestart Cortesi\surnameend} \&
  \bibinfo{author}{Martina \surnamestart Olliaro\surnameend}
  (\bibinfo{year}{2018}): \emph{\bibinfo{title}{M-String Segmentation: {A}
  Refined Abstract Domain for String Analysis in {C} Programs}}.
\newblock In: {\sl \bibinfo{booktitle}{{TASE}'18}}, pp. \bibinfo{pages}{1--8},
  \doi{10.1109/TASE.2018.00009}.

\bibitemdeclare{article}{costantini2015}
\bibitem{costantini2015}
\bibinfo{author}{G.~\surnamestart Costantini\surnameend},
  \bibinfo{author}{P.~\surnamestart Ferrara\surnameend} \&
  \bibinfo{author}{A.~\surnamestart Cortesi\surnameend} (\bibinfo{year}{2015}):
  \emph{\bibinfo{title}{A suite of abstract domains for static analysis of
  string values}}.
\newblock {\sl \bibinfo{journal}{Softw., Pract. Exper.}}
  \bibinfo{volume}{45}(\bibinfo{number}{2}), \doi{10.1002/spe.2218}.

\bibitemdeclare{inproceedings}{cousot1997}
\bibitem{cousot1997}
\bibinfo{author}{P.~\surnamestart Cousot\surnameend} (\bibinfo{year}{1997}):
  \emph{\bibinfo{title}{{T}ypes as {A}bstract {I}nterpretations}}.
\newblock In: {\sl \bibinfo{booktitle}{{POPL}'97}},
  \doi{10.1145/263699.263744}.

\bibitemdeclare{inproceedings}{cousot1977}
\bibitem{cousot1977}
\bibinfo{author}{P.~\surnamestart Cousot\surnameend} \&
  \bibinfo{author}{R.~\surnamestart Cousot\surnameend} (\bibinfo{year}{1977}):
  \emph{\bibinfo{title}{Abstract {I}nterpretation: {A} {U}nified {L}attice
  {M}odel for {S}tatic {A}nalysis of {P}rograms by {C}onstruction or
  {A}pproximation of {F}ixpoints}}.
\newblock In: {\sl \bibinfo{booktitle}{{POPL}'77}},
  \doi{10.1145/512950.512973}.

\bibitemdeclare{article}{cousot1992}
\bibitem{cousot1992}
\bibinfo{author}{P.~\surnamestart Cousot\surnameend} \&
  \bibinfo{author}{R.~\surnamestart Cousot\surnameend} (\bibinfo{year}{1992}):
  \emph{\bibinfo{title}{{A}bstract {I}nterpretation {F}rameworks}}.
\newblock {\sl \bibinfo{journal}{J. Log. Comput.}}
  \bibinfo{volume}{2}(\bibinfo{number}{4}), \doi{10.1093/logcom/2.4.511}.

\bibitemdeclare{inproceedings}{cousot1992-2}
\bibitem{cousot1992-2}
\bibinfo{author}{P.~\surnamestart Cousot\surnameend} \&
  \bibinfo{author}{R.~\surnamestart Cousot\surnameend} (\bibinfo{year}{1992}):
  \emph{\bibinfo{title}{Comparing the {G}alois {C}onnection and
  {W}idening/{N}arrowing {A}pproaches to {A}bstract {I}nterpretation}}.
\newblock In: {\sl \bibinfo{booktitle}{{PLILP}'92}},
  \doi{10.1007/3-540-55844-6\_142}.

\bibitemdeclare{inproceedings}{logozzo2011}
\bibitem{logozzo2011}
\bibinfo{author}{P.~\surnamestart Cousot\surnameend},
  \bibinfo{author}{R.~\surnamestart Cousot\surnameend} \&
  \bibinfo{author}{F.~\surnamestart Logozzo\surnameend} (\bibinfo{year}{2011}):
  \emph{\bibinfo{title}{A parametric segmentation functor for fully automatic
  and scalable array content analysis}}.
\newblock In: {\sl \bibinfo{booktitle}{{POPL}'11}}, pp.
  \bibinfo{pages}{105--118}, \doi{10.1145/1926385.1926399}.

\bibitemdeclare{inproceedings}{cousot1978}
\bibitem{cousot1978}
\bibinfo{author}{P.~\surnamestart Cousot\surnameend} \&
  \bibinfo{author}{N.~\surnamestart Halbwachs\surnameend}
  (\bibinfo{year}{1978}): \emph{\bibinfo{title}{Automatic {D}iscovery of
  {L}inear {R}estraints {A}mong {V}ariables of a {P}rogram}}.
\newblock In: {\sl \bibinfo{booktitle}{{POPL}'78}},
  \doi{10.1145/512760.512770}.

\bibitemdeclare{book}{davis1994}
\bibitem{davis1994}
\bibinfo{author}{M.~D. \surnamestart Davis\surnameend},
  \bibinfo{author}{R.~\surnamestart Sigal\surnameend} \& \bibinfo{author}{E.~J.
  \surnamestart Weyuker\surnameend} (\bibinfo{year}{1994}):
  \emph{\bibinfo{title}{Computability, {C}omplexity, and {L}anguages: Fund. of
  Theor. CS}}.
\newblock \bibinfo{publisher}{Academic Press Professional, Inc.},
  \doi{10.2307/2275691}.

\bibitemdeclare{inproceedings}{domaratzki2001}
\bibitem{domaratzki2001}
\bibinfo{author}{M.~\surnamestart Domaratzki\surnameend},
  \bibinfo{author}{J.~\surnamestart Shallit\surnameend} \&
  \bibinfo{author}{S.~\surnamestart Yu\surnameend} (\bibinfo{year}{2001}):
  \emph{\bibinfo{title}{Minimal {C}overs of {F}ormal {L}anguages}}.
\newblock In: {\sl \bibinfo{booktitle}{{DLT}'01}},
  \doi{10.1007/3-540-46011-X\_28}.

\bibitemdeclare{misc}{silva2006}
\bibitem{silva2006}
\bibinfo{author}{V.~\surnamestart {D'Silva}\surnameend} (\bibinfo{year}{2006}):
  \emph{\bibinfo{title}{Widening for {A}utomata}}.
\newblock \bibinfo{howpublished}{Diploma Thesis, Institut Fur Informatick,
  UZH}.

\bibitemdeclare{inproceedings}{fromherz18}
\bibitem{fromherz18}
\bibinfo{author}{A.~\surnamestart Fromherz\surnameend},
  \bibinfo{author}{A.~\surnamestart Ouadjaout\surnameend} \&
  \bibinfo{author}{A.~\surnamestart Min{\'{e}}\surnameend}
  (\bibinfo{year}{2018}): \emph{\bibinfo{title}{Static Value Analysis of Python
  Programs by Abstract Interpretation}}.
\newblock In: {\sl \bibinfo{booktitle}{{NFM}'18}},
  \doi{10.1007/978-3-319-77935-5\_14}.

\bibitemdeclare{article}{GM16}
\bibitem{GM16}
\bibinfo{author}{R.~\surnamestart Giacobazzi\surnameend} \&
  \bibinfo{author}{I.~\surnamestart Mastroeni\surnameend}
  (\bibinfo{year}{2016}): \emph{\bibinfo{title}{Making abstract models
  complete}}.
\newblock {\sl \bibinfo{journal}{MSCS}}
  \bibinfo{volume}{26}(\bibinfo{number}{4}), \doi{10.1017/S0960129514000358}.

\bibitemdeclare{inproceedings}{GQ01}
\bibitem{GQ01}
\bibinfo{author}{R.~\surnamestart Giacobazzi\surnameend} \&
  \bibinfo{author}{E.~\surnamestart Quintarelli\surnameend}
  (\bibinfo{year}{2001}): \emph{\bibinfo{title}{Incompleteness, counterexamples
  and refinements in abstract model-checking}}.
\newblock In: {\sl \bibinfo{booktitle}{{SAS}'01}},
  \doi{10.1007/3-540-47764-0\_20}.

\bibitemdeclare{article}{GRS00}
\bibitem{GRS00}
\bibinfo{author}{R.~\surnamestart Giacobazzi\surnameend},
  \bibinfo{author}{F.~\surnamestart Ranzato\surnameend} \&
  \bibinfo{author}{F.~\surnamestart Scozzari.\surnameend}
  (\bibinfo{year}{2000}): \emph{\bibinfo{title}{Making {A}bstract
  {I}nterpretation {C}omplete}}.
\newblock {\sl \bibinfo{journal}{{JACM}}}
  \bibinfo{volume}{47}(\bibinfo{number}{2}), \doi{10.1145/333979.333989}.

\bibitemdeclare{inproceedings}{hauzar2015}
\bibitem{hauzar2015}
\bibinfo{author}{D.~\surnamestart Hauzar\surnameend} \&
  \bibinfo{author}{J.~\surnamestart Kofron\surnameend} (\bibinfo{year}{2015}):
  \emph{\bibinfo{title}{{F}ramework for {S}tatic {A}nalysis of {PHP}
  {A}pplications}}.
\newblock In: {\sl \bibinfo{booktitle}{{ECOOP}'15}},
  \doi{10.4230/LIPIcs.ECOOP.2015.689}.

\bibitemdeclare{article}{afa}
\bibitem{afa}
\bibinfo{author}{L.~\surnamestart Hol{\'{\i}}k\surnameend},
  \bibinfo{author}{P.~\surnamestart Janku\surnameend},
  \bibinfo{author}{A.~\surnamestart Lin\surnameend},
  \bibinfo{author}{P.~\surnamestart R{\"{u}}mmer\surnameend} \&
  \bibinfo{author}{T.~\surnamestart Vojnar\surnameend} (\bibinfo{year}{2018}):
  \emph{\bibinfo{title}{String constraints with concatenation and transducers
  solved efficiently}}.
\newblock \doi{10.1145/3158092}.

\bibitemdeclare{book}{hopcroft1979}
\bibitem{hopcroft1979}
\bibinfo{author}{J.~\surnamestart Hopcroft\surnameend} \&
  \bibinfo{author}{J.~\surnamestart Ullman\surnameend} (\bibinfo{year}{1979}):
  \emph{\bibinfo{title}{{I}ntroduction to {A}utomata {T}heory, {L}anguages and
  {C}omputation}}.
\newblock \bibinfo{publisher}{Addison-Wesley}, \doi{10.1145/568438.568455}.

\bibitemdeclare{inproceedings}{tajs2009}
\bibitem{tajs2009}
\bibinfo{author}{S.~\surnamestart Jensen\surnameend},
  \bibinfo{author}{A.~\surnamestart M{\o}ller\surnameend} \&
  \bibinfo{author}{P.~\surnamestart Thiemann\surnameend}
  (\bibinfo{year}{2009}): \emph{\bibinfo{title}{{T}ype {A}nalysis for
  {J}ava{S}cript}}.
\newblock In: {\sl \bibinfo{booktitle}{{SAS}'09}},
  \doi{10.1007/978-3-642-03237-0\_17}.

\bibitemdeclare{inproceedings}{evil}
\bibitem{evil}
\bibinfo{author}{S.~H. \surnamestart Jensen\surnameend}, \bibinfo{author}{P.~A.
  \surnamestart Jonsson\surnameend} \& \bibinfo{author}{A.~\surnamestart
  M{\o}ller\surnameend} (\bibinfo{year}{2012}): \emph{\bibinfo{title}{Remedying
  the eval that men do}}.
\newblock In: {\sl \bibinfo{booktitle}{{ISSTA}'12}},
  \doi{10.1145/2338965.2336758}.

\bibitemdeclare{inproceedings}{jsai2014}
\bibitem{jsai2014}
\bibinfo{author}{V.~\surnamestart Kashyap\surnameend},
  \bibinfo{author}{K.~\surnamestart Dewey\surnameend},
  \bibinfo{author}{E.~\surnamestart Kuefner\surnameend},
  \bibinfo{author}{J.~\surnamestart Wagner\surnameend},
  \bibinfo{author}{K.~\surnamestart Gibbons\surnameend},
  \bibinfo{author}{J.~\surnamestart Sarracino\surnameend},
  \bibinfo{author}{B.~\surnamestart Wiedermann\surnameend} \&
  \bibinfo{author}{B.~\surnamestart Hardekopf\surnameend}
  (\bibinfo{year}{2014}): \emph{\bibinfo{title}{{JSAI:} a static analysis
  platform for {J}ava{S}cript}}.
\newblock In: {\sl \bibinfo{booktitle}{{FSE}'14}},
  \doi{10.1145/2635868.2635904}.

\bibitemdeclare{inproceedings}{safe2012}
\bibitem{safe2012}
\bibinfo{author}{H.~\surnamestart Lee\surnameend},
  \bibinfo{author}{S.~\surnamestart Won\surnameend},
  \bibinfo{author}{J.~\surnamestart Jin\surnameend},
  \bibinfo{author}{J.~\surnamestart Cho\surnameend} \&
  \bibinfo{author}{S.~\surnamestart Ryu\surnameend} (\bibinfo{year}{2012}):
  \emph{\bibinfo{title}{{SAFE}: {F}ormal specification and implementation of a
  scalable analysis framework for {ECMAS}cript}}.
\newblock In: {\sl \bibinfo{booktitle}{{FOOL}'12}}.

\bibitemdeclare{inproceedings}{popl2016}
\bibitem{popl2016}
\bibinfo{author}{A.~Widjaja \surnamestart Lin\surnameend} \&
  \bibinfo{author}{P.~\surnamestart Barcel{\'{o}}\surnameend}
  (\bibinfo{year}{2016}): \emph{\bibinfo{title}{String solving with word
  equations and transducers: towards a logic for analysing mutation {XSS}}}.
\newblock In: {\sl \bibinfo{booktitle}{{POPL}'16}},
  \doi{10.1145/2837614.2837641}.

\bibitemdeclare{inproceedings}{midtgaard16}
\bibitem{midtgaard16}
\bibinfo{author}{J.~\surnamestart Midtgaard\surnameend},
  \bibinfo{author}{F.~\surnamestart Nielson\surnameend} \&
  \bibinfo{author}{H.~R. \surnamestart Nielson\surnameend}
  (\bibinfo{year}{2016}): \emph{\bibinfo{title}{A {P}arametric {A}bstract
  {D}omain for {L}attice-{V}alued {R}egular {E}xpressions}}.
\newblock In: {\sl \bibinfo{booktitle}{{SAS}'16}},
  \doi{10.1007/978-3-662-53413-7\_17}.

\bibitemdeclare{inproceedings}{park2016}
\bibitem{park2016}
\bibinfo{author}{C.~\surnamestart Park\surnameend},
  \bibinfo{author}{H.~\surnamestart Im\surnameend} \&
  \bibinfo{author}{S.~\surnamestart Ryu\surnameend} (\bibinfo{year}{2016}):
  \emph{\bibinfo{title}{Precise and scalable static analysis of {jQuery} using
  a regular expression domain}}.
\newblock In: {\sl \bibinfo{booktitle}{{DLS}'16}},
  \doi{10.1145/2989225.2989228}.

\bibitemdeclare{inproceedings}{safe}
\bibitem{safe}
\bibinfo{author}{C.~\surnamestart Park\surnameend} \&
  \bibinfo{author}{S.~\surnamestart Ryu\surnameend} (\bibinfo{year}{2015}):
  \emph{\bibinfo{title}{{S}calable and {P}recise {S}tatic {A}nalysis of
  {J}ava{S}cript {A}pplications via {L}oop-{S}ensitivity}}.
\newblock In: {\sl \bibinfo{booktitle}{{ECOOP}'15}},
  \doi{10.4230/LIPIcs.ECOOP.2015.735}.

\bibitemdeclare{inproceedings}{pradel2015}
\bibitem{pradel2015}
\bibinfo{author}{M.~\surnamestart Pradel\surnameend} \&
  \bibinfo{author}{K.~\surnamestart Sen\surnameend} (\bibinfo{year}{2015}):
  \emph{\bibinfo{title}{The {G}ood, the {B}ad, and the {U}gly: {A}n {E}mpirical
  {S}tudy of {I}mplicit {T}ype {C}onversions in {J}ava{S}cript}}.
\newblock In: {\sl \bibinfo{booktitle}{{ECOOP}'15}},
  \doi{10.4230/LIPIcs.ECOOP.2015.519}.

\bibitemdeclare{inproceedings}{pribavkina2010}
\bibitem{pribavkina2010}
\bibinfo{author}{E.~\surnamestart Pribavkina\surnameend} \&
  \bibinfo{author}{E.~\surnamestart Rodaro\surnameend} (\bibinfo{year}{2010}):
  \emph{\bibinfo{title}{{S}tate {C}omplexity of {P}refix, {S}uffix, {B}ifix and
  {I}nfix {O}perators on {R}egular {L}anguages}}.
\newblock In: {\sl \bibinfo{booktitle}{{DLT}'10}},
  \doi{10.1007/978-3-642-14455-4\_34}.

\bibitemdeclare{book}{cfa}
\bibitem{cfa}
\bibinfo{author}{M~\surnamestart Sharir\surnameend} \&
  \bibinfo{author}{A~\surnamestart Pnueli\surnameend} (\bibinfo{year}{1978}):
  \emph{\bibinfo{title}{{Two approaches to interprocedural data flow
  analysis}}}.
\newblock \bibinfo{publisher}{NYU CS}, \bibinfo{address}{NY}.

\bibitemdeclare{misc}{w3school-string}
\bibitem{w3school-string}
\bibinfo{author}{\surnamestart W3S\surnameend}: \emph{\bibinfo{title}{{{JS}
  {S}tring {R}ef.}}}
\newblock
  \bibinfo{howpublished}{\url{www.w3schools.com/jsref/jsref_obj_string.asp}}.
\newblock \bibinfo{note}{Accessed 16-06-2018}.

\bibitemdeclare{inproceedings}{xu2012}
\bibitem{xu2012}
\bibinfo{author}{W.~\surnamestart Xu\surnameend},
  \bibinfo{author}{F.~\surnamestart Zhang\surnameend} \&
  \bibinfo{author}{S.~\surnamestart Zhu\surnameend} (\bibinfo{year}{2012}):
  \emph{\bibinfo{title}{The power of obfuscation techniques in malicious
  {J}ava{S}cript code: {A} measurement study}}.
\newblock In: {\sl \bibinfo{booktitle}{{MALWARE}'12}},
  \doi{10.1109/MALWARE.2012.6461002}.

\bibitemdeclare{inproceedings}{yu2008}
\bibitem{yu2008}
\bibinfo{author}{F.~\surnamestart Yu\surnameend},
  \bibinfo{author}{T.~\surnamestart Bultan\surnameend},
  \bibinfo{author}{M.~\surnamestart Cova\surnameend} \& \bibinfo{author}{O.~H.
  \surnamestart Ibarra\surnameend} (\bibinfo{year}{2008}):
  \emph{\bibinfo{title}{{S}ymbolic {S}tring {V}erification: {A}n
  {A}utomata-{B}ased {A}pproach}}.
\newblock In: {\sl \bibinfo{booktitle}{{SPIN}'08}},
  \doi{10.1007/978-3-540-85114-1\_21}.

\bibitemdeclare{article}{YuZS94}
\bibitem{YuZS94}
\bibinfo{author}{S.~\surnamestart Yu\surnameend},
  \bibinfo{author}{Q.~\surnamestart Zhuang\surnameend} \&
  \bibinfo{author}{K.~\surnamestart Salomaa\surnameend} (\bibinfo{year}{1994}):
  \emph{\bibinfo{title}{The {S}tate {C}omplexities of {S}ome {B}asic
  {O}perations on {R}egular {L}anguages}}.
\newblock {\sl \bibinfo{journal}{Theor. Comput. Sci.}}
  \bibinfo{volume}{125}(\bibinfo{number}{2}),
  \doi{10.1016/0304-3975(92)00011-F}.

\end{thebibliography}

%\newpage

%\appendix
%\input{app}
\end{document}